\documentclass[aps,prc,twocolumn,preprintnumbers,amsmath,amssymb,showkeys,floatfix,nofootinbib,reprint]{revtex4-1}

\usepackage{amsmath}
\usepackage{amsfonts}
\usepackage{amssymb}
\usepackage{dcolumn}

\usepackage[toc,page]{appendix}
\usepackage{graphicx}
\usepackage{siunitx}
\usepackage{booktabs}
\usepackage{longtable}
\usepackage[version=4]{mhchem}
\usepackage{color}
\usepackage{bm}
\usepackage{hyperref}
\usepackage{diagbox} 

\usepackage{comment}

\newcolumntype{P}[1]{>{\centering\arraybackslash}p{#1}}

\begin{document}

\title{Exploring laser-driven neutron sources for neutron capture cascades and the production of neutron-rich isotopes}

\author{Paul Hill}
\affiliation{Max-Planck-Institut f\"ur Kernphysik, Saupfercheckweg 1, D-69117 Heidelberg, Germany}

\author{Yuanbin Wu}
\email[Corresponding author. ]{yuanbin.wu@mpi-hd.mpg.de}
\affiliation{Max-Planck-Institut f\"ur Kernphysik, Saupfercheckweg 1, D-69117 Heidelberg, Germany}

\begin{abstract}
The production of neutron-rich isotopes and the occurrence of neutron capture cascades via laser-driven (pulsed) neutron sources are investigated theoretically. The considered scenario involves the interaction of a laser-driven neutron beam with a target made of a single type of seed nuclide. We present a comprehensive study over $95$ seed nuclides in the range $3\le Z \le 100$ from $^7_3$Li to $^{255}_{100}$Fm. For each element, the heaviest sufficiently-long-lived (half life $> 1$ h) isotope whose data is available in the recent ENDF-B-VIII.0 neutron sublibrary is considered. We identify interesting seed nuclides with good performance in the production of neutron-rich isotopes where neutron capture cascades may occur. The effects of the neutron number per pulse, the neutron-target interaction size and the number of neutron pulses are also analyzed. Our results show the possibility of observing up to $4$ successive neutron capture events leading to neutron-rich isotopes with $4$ more neutrons than the original seed nuclide. This hints at new experimental possibilities to produce neutron-rich isotopes and simulate neutron capture nucleosynthesis in the laboratory. With several selected interesting seed nuclides in the region of the branching point of the $s$-process ($^{126}_{51}$Sb, $^{176}_{71}$Lu and $^{187}_{75}$Re) or the waiting point of the $r$-process (Lu, Re, Os, Tm, Ir and Au), we expect that laser-driven experiments can shed light on our understanding of nucleosynthesis.
\end{abstract}

\date{\today}

\maketitle

%%----------------------------------------------------------introduction-----------------------------------------------------------

\section{Introduction}

Neutron-rich isotopes are of particular interest in both fundamental nuclear physics \cite{ThoennessenRPP2013, AhnPRL2019, GorgesPRL2019, ZhangPRL2019, CrawfordPRL2019, GatesPRL2018, TarasovPRL2018} and the neutron capture processes of nucleosynthesis in astrophysics \cite{BurbidgeRMP1957, KappelerRMP2011, ArnouldPR2007}. On the fundamental nuclear physics side, neutron-rich isotopes could provide key information to test nuclear models and to understand the nuclear interaction \cite{ThoennessenRPP2013, AhnPRL2019, GorgesPRL2019, ZhangPRL2019, CrawfordPRL2019, GatesPRL2018, TarasovPRL2018}. On the astrophysics side, the slow ($s$-process) and the rapid neutron capture process ($r$-process) of neutron capture nucleosynthesis contribute in roughly equal amount to the total elemental abundances beyond iron \cite{KappelerRMP2011, ArnouldPR2007, MeyerARAA1994, CowanPT2004, SnedenS2003}. Although the neutron capture nucleosynthesis has been studied extensively, some issues still remain open, such as the $s$-process branching \cite{KappelerRMP2011, WallersteinRMP1997} and the astrophysical sites of the $r$-process \cite{ArnouldPR2007, WallersteinRMP1997, PanovAA2009, FreiburghausAPJL1999, SurmanAPJL2008, ThielemannARNPS2017, WanajoAPJ2018, RosswogAA2018, FrebelARNPS2018, HorowitzJPG2019, ShibataARNPS2019, KajinoPPNP2019, ArnouldPPNP2020, RadiceARNPS2020}. Recently, observations connected to the neutron star merger event GW170817 \cite{AbbottPRL2017} have confirmed the emission of a kilonova afterglow which would be powered by the radioactive decay of heavy nuclides produced in the $r$-process \cite{ArcaviNature2017, PianNature2017, SmarttNature2017, KasenNature2017, Soares-SantosAPJL2017, CowperthwaiteAPJL2017, NichollAPJL2017, ChornockAPJL2017, ValentiAPJL2017, TanvirAPJL2017, VillarAPJL2017, CoulterScience2017, KasliwalScience2017, DroutScience2017, ShappeeScience2017, KilpatrickScience2017}, leading to a rapidly increased attention on the topic \cite{WanajoAPJ2018, RosswogAA2018, FrebelARNPS2018, HorowitzJPG2019, ShibataARNPS2019, KajinoPPNP2019, ArnouldPPNP2020, RadiceARNPS2020}. Sensitivity studies \cite{MumpowerPPNP2016} have also shown that significant variations of the final abundances in the neutron capture nucleosynthesis due to the uncertainties of the properties of relevant neutron-rich nuclei. Measurements of the properties of neutron-rich nuclei on or near the $s$-process and $r$-process paths will improve our understanding of the neutron capture nucleosynthesis \cite{KappelerRMP2011, ArnouldPR2007, rproc, NegoitaRRP2016, HabsAPB2011}. Furthermore, radioisotopes also have extensive applications in industry \cite{CharltonBook1986} and medicine \cite{radmed}, as well as in the study of nuclear batteries considered as potential long-lived small-scale power sources \cite{DuggiralaBook2010, PrelasBook2016}.

So far, the production of neutron-rich isotopes mainly relies on accelerator- and reactor-based facilities, via neutron capture, projectile fragmentation, projectile fission or nuclear fusion reactions \cite{ThoennessenRPP2013, AhnPRL2019, GorgesPRL2019, ZhangPRL2019, CrawfordPRL2019, GatesPRL2018, TarasovPRL2018, nucchart, rproc, radreac, ThoennessenBook2016}. However, the development of laser technology in the past decades provides a powerful tool for the study of nuclear physics and nuclear astrophysics in laser-generated plasmas \cite{NegoitaRRP2016, WuAPJ2017, CaseyPRL2012, ZylstraPRL2016, BleuelPFR2016, NIF2018}. High-power lasers offer the opportunity of generating intense neutron beams at comparatively small-scale laser facilities, allowing for the production of neutron-rich isotopes via neutron capture. With such lasers, the neutron beams can be produced via laser-induced particle acceleration leading to high-energy particle beams that subsequently interact with a secondary target (beam-target interaction) \cite{nsrcrit, PomerantzPRL2014, HigginsonPRL2015}, or via thermonuclear reactions \cite{nsrcwu, RenPRL2017, MaPRL2015, DoppnerPRL2015, OlsonPRL2016}. While they provide a comparatively low number of neutrons per pulse, the achievable neutron fluxes can be a few orders of magnitude higher than the ones in the conventional neutron sources at large-scale rector- and accelerator-based facilities \cite{PomerantzPRL2014, PomerantzELIS2015, VogelJOM2012, CarlileBook2016}. With the high-power lasers (short pulse duration Petawatt-class lasers, or nanosecond duration kilo-Joule or Mega-Joule lasers) available nowadays or in the near future, intense pulses of $10^{10}$ neutrons or higher within durations on the order of picoseconds or nanoseconds can be obtained \cite{nsrcrit, PomerantzPRL2014, HigginsonPRL2015, KarNJP2016, MirfayziAPL2017, KleinschmidtPOP2018, nsrcwu, RenPRL2017, MaPRL2015, DoppnerPRL2015, OlsonPRL2016}, and consequently very high neutron fluxes on the level of $10^{20}$ n/cm$^2$/s are achievable. With the intense laser-driven neutron beams, neutron capture experiments and the production of neutron-rich isotopes via neutron capture become possible \cite{PomerantzPRL2014, PomerantzELIS2015, NIF2018} at such small-scale laser facilities.  As achievable neutron fluxes are that high, another advantage of studying neutron capture processes with laser-driven neutron sources is that they provide an opportunity to analyze neutron capture cascades similar to the ones occurring during neutron capture nucleosynthesis in astrophysics. This would give us the chance to simulate the neutron capture nucleosynthesis in the laboratory for the first time \cite{PomerantzPRL2014, PomerantzELIS2015, NIF2018}, leading to an improved understanding of the ongoing processes.

In this paper, we conduct a comprehensive study of the production of neutron-rich isotopes and neutron capture cascades taking place in single-component targets being irradiated by a laser-driven (pulsed) neutron source. A comprehensive study over $95$ potential seed nuclides in the range $3\le Z \le 100$ from $^7_3$Li to $^{255}_{100}$Fm is conducted. For each element, the heaviest sufficiently-long-lived (half life $> 1$ h) isotope whose data is available in the recent ENDF-B-VIII.0 neutron sublibrary \cite{ENDFBVIII0} is considered as a potential seed nuclide. We are interested in the seed nuclides which are in the region of the branching point of the $s$-process or the waiting point of the $r$-process. We are also interested in the production of neutron-rich isotopes in a regime which has never been accessed by other means in the laboratory. 

Our calculations are performed by a one-dimensional (1D) model, accounting for the successive radiative neutron capture process, the damping of the incident neutron beam, and the loss of nuclei by transmutation and radioactive decay. Both single and multi neutron pulse scenarios are analysed. Laser-driven (pulsed) neutron beams with average energy in the range between $50$ keV and $10$ MeV are considered. According to our numerical results, interesting seed nuclides are identified from the large list of potential seed nuclides, with good performance in the production of neutron-rich isotopes where successive neutron capture process may occur. For a scenario of $10^4$ shots, from a laser-driven neutron source generating $10^{12}$ neutrons per pulse ($10^{12}$ n/pl) at a repetition rate of $1$ Hz, our results show the possibility of observing up to $4$ successive neutron capture events leading to the production of neutron-rich isotopes with $4$ more neutrons than the original seed nuclide. This provides the chance of simulating the astrophysical neutron capture nucleosynthesis in the laboratory. Among the identified interesting seed nuclides, some are in the region of the branching point of the $s$-process ($^{126}_{51}$Sb, $^{176}_{71}$Lu and $^{187}_{75}$Re) or the waiting point of the $r$-process (Lu, Re, Os, Tm, Ir and Au). It is also possible to produce neutron-rich isotopes ($^{248}_{95}$Am, $^{258}_{99}$Es and $^{259}_{99}$Es) in a regime that has not been accessed in the laboratory. We note that such intense laser-driven neutron beams with a high repetition rate are expected to be achievable in the Petawatt-class laser facilities available in the near future \cite{NegoitaRRP2016, nsrcrit, PomerantzPRL2014, HigginsonPRL2015, nsrcwu, PomerantzELIS2015}, such as the Extreme Light Infrastructure (ELI) facilities which are under construction \cite{NegoitaRRP2016, ELIweb2020}.

The paper is organized as follows. In Sec.~\ref{sec:ths}, we present the theoretical approach used to compute neutron captures in a target being irradiated by a laser-driven neutron beam. In Sec.~\ref{sec:seeddata}, we present the target configuration and the potential seed nuclides as well as the data sources for the reaction cross sections. Our numerical results and discussions are presented in Sec.~\ref{sec:nrone} for the case of one neutron pulse and in Sec.~\ref{sec:nrmul} for the case of multiple neutron pulses. We finally summarize and conclude in Sec.~\ref{sec:sum}.

%%----------------------------------------------------------theory-----------------------------------------------------------

\section{Theoretical approach \label{sec:ths}}

We investigate a setup in which a rectangular target made of a pure seed material is irradiated by a laser-driven neutron beam. The number of neutrons per pulse of the neutron beam is denoted by $N_p$. Average neutron energies in the range between $50$ keV and $10$ MeV are considered, which cover the range of the neutron energies of the laser-driven neutron sources \cite{nsrcrit, PomerantzPRL2014, HigginsonPRL2015, KarNJP2016, MirfayziAPL2017, KleinschmidtPOP2018, nsrcwu, RenPRL2017, MaPRL2015, DoppnerPRL2015, OlsonPRL2016} for the purpose of the present study. The target has an interacting surface area $A$ hit by the neutron beam and a thickness $L$. The successive radiative neutron capture process producing neutron-enriched nuclei, the damping of the incident neutron beam, and the loss of nuclei by transmutation and radioactive decay are of interest and have been taken into account. We note that it is a one-dimensional model, i.e., the neutron-target interaction happens only in the volume $A \times L$ and is homogeneous in the interacting area $A$.

Inside the target, beam neutrons interact with the target atoms generating secondary particles, thereby also transmuting the target nuclides. Successive effects of these secondary particles are neglected for the purpose of the present study. Furthermore, as we are not interested in the production of nuclides other than the neutron-enriched isotopes of the seed nuclide, we do not calculate abundances or successive transmutations of such nuclides, i.e., we only keep track of the seed nuclide and the neutron-enriched isotopes of the seed nuclide. By doing so, we neglect the contribution of any loops in a transmutation path (e.g., $\ce{^28Si ->[($n$, $p$)] ^28Al ->[\beta^-] ^28Si}$), whose effect is assumed to be small. In order to close a loop, processes that kick out a nucleon from the nucleus have to take place. However, such processes like ($n$, $p$), ($n$, $2n$), ($n$, $d$), etc., are in general highly suppressed below several MeV and could therefore only skew our results towards the high energy end. We note that more sophisticated approaches modelling a variety of different processes, including such loops, can be found in, e.g., Refs.~\cite{WilsonPHD1999, KumNET2018, SubletNDS2017}.

In the following of Sec.~\ref{sec:ths}, we begin with the simplest case that the neutron capture in a thin target during one neutron pulse in Sec.~\ref{sec:ththin}. Then we generalise the model to the case of a thick target in Sec.~\ref{sec:ththick} to include the effect of the damping of the incident neutron beam. In Sec.~\ref{sec:thmul}, we further generalise the model to the case of multiple neutron pulses to include the effects of the multiple neutron pulses and the decay of the nuclei. In Sec.~\ref{sec:thtot}, we calculate the total amount of neutron-enriched nuclei produced during the interaction, which in addition to the number of nuclei remaining after the interaction also accounts for the number of nuclei lost due to transmutation and radioactive decay.

\subsection{Neutron capture in a thin target during one neutron pulse \label{sec:ththin}}

We begin with the case of a thin target with thickness $L\ll \lambda$ and further assume that the interaction time $T_p$ (the neutron pulse duration) is much shorter than all half lives of the involved particles. Here, $\lambda$ is the neutron penetration depth of the target material, which marks the scale at which the damping of the incident neutron beam becomes relevant. Furthermore only one neutron pulse is considered for the moment. Based on these assumptions, we can neglect the damping of the incident neutron beam and the decay of the nuclei. We denote the isotope having $i$ more neutrons than the seed isotope ($0$-species) as the $i$-species isotope. The populations of these isotopes are coupled via the following set of equations \cite{SubletNDS2017}
\begin{align}
\dot{N_0} &= - \sigma_{tr,0}R_b\frac{N_0(t)}{A} \label{eq:thinNz}\\
\dot{N_1} &= - \sigma_{tr,1}R_b\frac{N_1(t)}{A} + \sigma_{c,0}R_b\frac{N_0(t)}{A} \\
&\;\;\vdots \nonumber \\
\begin{split}
\dot{N_l} &= - \sigma_{tr,l}R_b\frac{N_l(t)}{A} + \sigma_{c,l-1}R_b\frac{N_{l-1}(t)}{A}\\
&\approx \sigma_{c,l-1}R_b\frac{N_{l-1}(t)}{A}
\end{split}\\
\dot{N_i} &\approx 0, i> l, \label{eq:thinNl}
\end{align}
where $N_i$ stands for the number of nuclei of the $i$-species isotope involved in the interaction, which initially takes on the value $N_i = N_i^0$. Here, the $l$-species is the cut off species that $N_l$ stays sufficiently small such that $N_i$ for $i>l$ is negligible. $R_b$ is the rate of neutrons irradiating the target, which is related to the neutron current density $j_b$ via $R_b = j_b A$. Moreover, $\sigma_{c,i}$ denotes the neutron capture cross section of the $i$-species isotope, and $\sigma_{tr,i}$ is its transmutation cross section including all processes changing the neutron or proton number. We note that the cross sections need to be averaged out (`collapsed') \cite{SubletNDS2017} by the flux spectrum weighting into energy independent values, when accounting the neutron beam has an energy spectrum.

The beam rate $R_b$ in Eqs.~(\ref{eq:thinNz})-(\ref{eq:thinNl}) is in general time dependent. However, this time dependence can be eliminated by changing $t \rightarrow \tau(t)=\int_0^tR_b(t')dt'/N_p$, which effectively replaces $R_b(t)$ with $N_p$ in the above equations. Eventually, we are only interested in the population at the end of the pulse at $T_p$ [$\tau (T_p) = 1$]. For simplicity we just rename $\tau \rightarrow t$ again and evaluate at $t=1$. We further define the capture and loss parameters 
\begin{equation}
\mu_i = N_p \sigma_{c,i}/A,\quad\eta_i = N_p \sigma_{tr,i}/A
\end{equation}
for the $i$-species isotope, whose values can be estimated for the purpose of this study as $\mu_i, \eta_i \lessapprox 10^{-6}-10^{-4}$.

The above equations (\ref{eq:thinNz})-(\ref{eq:thinNl}) can be rewritten in a more convenient way as
\begin{align}
\bm{\dot{N}} &= \bm{B}\bm{N},
\end{align}
where the matrix 
\begin{equation}
\bm{B} = \left(\begin{matrix}-\eta_0&0&0&\cdots&0\\\mu_0&-\eta_1&0&\cdots&0\\0&\mu_1&-\eta_2&\cdots&0\\\vdots&\vdots&\vdots&\ddots&\vdots\\0&\cdots&0&\mu_{l-1}&0\end{matrix}\right)
\end{equation}
has been introduced and $\bm{N} = \left(\begin{matrix} N_0 & N_1 & \cdots & N_l \end{matrix}\right)^T$.

The solution can be directly obtained by integration, yielding $\bm{N}(t) = e^{\bm{B}t}\bm{N_0}$ \cite{JeffreyBook2010}. Hence we find the populations after one pulse \cite{JeffreyBook2010},
\begin{equation}
\bm{N}(1\text{pl}) = e^{\bm{B}}\bm{N}_0 = \sum_{k=0}^\infty\frac{\bm{B}^k}{k!} \bm{N}_0, \label{eq:1pl}
\end{equation}
where $\bm{N}_0 = \left(\begin{matrix} N_0^0 & N_1^0 & \cdots & N_l^0 \end{matrix}\right)^T$ is the initial value of $\bm{N}$.

\subsection{Neutron capture in a thick target during one neutron pulse \label{sec:ththick}}

Let us now relax the assumption of a thin target and account for the damping of the incident neutron beam \cite{FlemingIJARI1982}. Since the initial seed nuclides will stay the most abundant species, the incident neutron beam is damped in good approximation according to
\begin{equation}
R_b(x) = R_{b,0}e^{-n_t \sigma_{tot,0} x},
\end{equation}
where $R_{b,0}$ is the initial neutron rate,  $n_t$ is the initial number density of the nuclei in the target and $\sigma_{tot,0}$ is the total neutron interaction cross section of the seed nuclide. The penetration depth $\lambda$ is then given by
\begin{equation} \label{eq:pend}
\lambda = \frac{1}{n_t \sigma_{tot,0}}.
\end{equation}
We can generalise the result in Sec.~\ref{sec:ththin} to this case by replacing $N_i\rightarrow dN_i$, $N_i^0 \rightarrow dN_i^0 = \frac{N_i^0}{L}dx$  and $\mu_i \rightarrow \mu_i(x) = N_p(x) \sigma_{c,i}/A$ (analogously for $\eta_i$), with $N_p(x) = N_pe^{-x/\lambda}$. The $e^{-x/\lambda}$ factor can be pulled out of $\bm{B}$, such that one effectively has $\bm{B}\rightarrow e^{-x/\lambda}\bm{B}$.

Integrating over $x$ can be performed in the sum representation of Eq.~(\ref{eq:1pl}) and we obtain
\begin{equation}
\bm{N}(1\text{pl}) = \sum_{k=0}\frac{\gamma_k \bm{B}^k}{k!}\bm{N}_0 \equiv \bm{M}\bm{N}_0, \label{eq:1pl_ex}
\end{equation}
where in practice we evaluate $\bm{M}$ up to $l$-th order, and define $\gamma_k = \lambda \left(1-e^{-kL/\lambda}\right)/(k L)$. We note that Eq.~(\ref{eq:1pl_ex}) is obtained under the assumption of spatially homogeneous in the target density \cite{FlemingIJARI1982}.

\subsection{Neutron capture during multiple neutron pulses \label{sec:thmul}}

So far we have only considered a single pulse. Let us now turn to the case that the neutron source repeatedly generates neutron pulses with repetition rate $f_{rep}$, i.e., each $T_{del} = 1/f_{rep}$ there is a neutron pulse. To describe the populations after several neutron pulses, radioactive decay of the nuclei also needs to be taken into account. Since we have assumed that all mean lifetimes $\tau_i$ are much longer than the neutron pulse duration $T_p$ ($T_p \ll T_{del}$), the model can be easily generalized. The populations after $s$ neutron pulses [plus ($T_{del} - T_p$) more precisely, i.e., at the time of $s T_{del}$] $\bm{N}(s)$ can be related to the populations after the previous pulse $\bm{N}(s-1)$ via \cite{SpanglerFED1993}
\begin{equation}
N_i(s) = e^{-T_{del}/\tau_i}\left[\bm{M}\bm{N}(s-1)\right]_i,\label{eq:nik}
\end{equation}
where the prefactor accounts for the radioactive decay, and the term $\bm{M}\bm{N}(s-1)$ gives the one-pulse result for the initial conditions $\bm{N}(s-1)$ (with the subscript $i$ on a vector, we denote the $i$-th component of the corresponding vector). With Eq.~(\ref{eq:nik}) the populations can be computed recursively.

In order to have some qualitative insights into the effect of multiple neutron pulses and Eq.~(\ref{eq:nik}), we assume $\tau_i = \infty$ and $\gamma_k = 1$ (i.e., thin target assumption). Then Eq.~(\ref{eq:nik}) becomes
\begin{equation}
\bm{N}(s) = \bm{M}^s\bm{N}_0 = e^{s\bm{B}}\bm{N}_0 = \sum_k \frac{s^k \bm{B}^k}{k!} \bm{N}_0.
\end{equation}
If $s\mu_i\ll 1$ and $s\eta_i\ll 1$ we may keep terms only to leading order in $k$ for each component of $\bm{N}(s)$. With the initial condition $\bm{N}_0 = \left(\begin{matrix} N_t & 0 & 0 & \cdots \end{matrix}\right)^T$, where $N_t$ is the initial number of seed nuclei in the neutron-target interaction region, the leading order for the $i$-species is the $\bm{B}^i$ term and we find that the populations scale polynomially with $s$ (the number of neutron pulses $N_{pl}$, respectively) as
\begin{equation}
N_i(s) = \frac{s^i}{i!}\left(\prod_{m=0}^{i-1}\mu_m \right) N_t, ~i>0. \label{eq:N_scaling}
\end{equation}
Furthermore the number of nuclei of the isotope that captured one more neutron is suppressed by a factor
\begin{equation}
\frac{N_{i+1}(s)}{N_i(s)} = \frac{s\mu_i}{i+1}, ~i>0,
\end{equation}
which is linear in $s$.

\subsection{Total amount of produced nuclides \label{sec:thtot}}

A further quantity of interest is the total amount of nuclei of the different species $N^\text{tot}_i$ produced during the interaction, which is governed by the differential equation
\begin{equation} \label{eq:toteq}
\frac{dN^{\text{tot}}_{i+1}}{dt} = \mu_iN_i(t). 
\end{equation}
Plugging in the in-pulse-solution $\bm{N}(t) = \sum_{k=0}\frac{\gamma_kB^kt^k}{k!}\bm{N}_0$ and integrating Eq.~(\ref{eq:toteq}) yields 
\begin{align}
\begin{split}
\Delta N^\text{tot}_{i+1}(1\text{pl}) &= \mu_i\left(\sum_{k=0}\frac{\gamma_kB^k}{(k+1)!}\bm{N}_0\right)_i \\
&\equiv \mu_i\left(\bm{M}'\bm{N}_0\right)_i,
\end{split}
\end{align}
for the total amount of nuclei of the $(i+1)$-species isotope produced during one neutron pulse. Summing up the contributions for $s$ neutron pulses gives
\begin{align}
\begin{split}
N^\text{tot}_{i+1}(s) &= \mu_i \sum_j^{s-1} \left[\bm{M}' \bm{N}(j) \right]_i \\
&= \mu_i \left[\bm{M}' \sum_{j=0}^{s-1} \bm{N}(j) \right]_i. \label{eq:ntik}
\end{split}
\end{align}

%%----------------------------------------------------------target and data-----------------------------------------------------------

\section{Seed nuclides and data sources \label{sec:seeddata}}

\subsection{Seed nuclides \label{sec:seednuc}}

Laser-driven neutron sources generate neutron pulses on the scale of micrometers with a pulse duration on the order of picoseconds or nanoseconds ($T_p \sim$ picoseconds to nanoseconds) \cite{nsrcrit, PomerantzPRL2014, HigginsonPRL2015, KarNJP2016, MirfayziAPL2017, KleinschmidtPOP2018, nsrcwu, RenPRL2017, MaPRL2015, DoppnerPRL2015, OlsonPRL2016}, and a repetition rate of the laser repetition rate (we assume in the present work $f_{rep} = 1$ Hz). Thus, for the  purpose of the present work, we assume a rectangular target with an interacting surface area $A = 25$ $\mu$m (perpendicular to the incident beam) and thickness $L = 100$ $\mu$m. Neutron beams with average energy $E_{inc}$ in the range between $50$ keV and $10$ MeV are considered, and a Gaussian profile energy spectrum with relative width of $w/E_{inc} = 10\%$ is assumed. We note that, according to Eqs.~(\ref{eq:thinNz})-(\ref{eq:thinNl}) and the discussion in Sec.~\ref{sec:ththin}, the production of neutron-rich isotopes depends on the incident neutron beam via the parameter $N_p/A$. Thus, assuming a fixed $L$ and a given neutron beam energy, $N_i/A$, where $N_i$ is a solution of Eqs.~(\ref{eq:thinNz})-(\ref{eq:thinNl}), is a function of $N_p/A$. This relation only holds for $N_i/A$, since the initial conditions of $N_i$ depend linearly on $A$ and $N_i$ itself depends linearly on these initial conditions. In the present work, we mainly focus on the case of $N_p = 10^{12}$ n/pl. Assuming the neutron pulse duration to be on the order of nanoseconds, this corresponds to a neutron flux of $\sim 10^{19}$ n/cm$^2$/s for laser-driven neutron sources. We note that highest neutron flux from conventional neutron sources under construction is expected to be around $10^{18}$ n/cm$^2$/s in European Spallation Source ESS, while most of the neutron fluxes from conventional neutron sources at large-scale rector- and accelerator-based facilities in operation or under construction are more than $3$ orders of magnitude lower than that from ESS \cite{CarlileBook2016}. This leads to at least $i$ orders of magnitude less for the production of the $i$-species isotope at the energies concerned, when comparing our result for laser-driven neutron sources to conventional neutron sources at large-scale rector- and accelerator-based facilities.

The main cross section data employed in the present work is the recent Evaluated Nuclear Data File (ENDF) release ENDF/B-VIII.0 by the National Nuclear Data Center (NNDC) from 2018 \cite{ENDFBVIII0}. We consider the heaviest sufficiently-long-lived (half life $> 1$ h) isotope per element whose data is available in ENDF/B-VIII.0 neutron sublibrary to serve as a potential seed target material. In addition, the NON-SMOKER library \cite{RAUSCHER200147} which provides theoretical neutron-capture cross sections is used for neutron-enriched isotopes.

A list of the seed nuclides considered in the present study is shown in Table~\ref{tab:info}, together with the half lives $T_{1/2}$ and $T_{1/2}^{+i}$ ($i$: $1$, $2$, $3$, and $4$) of each seed nuclide and its $i$-species neutron-enriched isotope, respectively. Furthermore, we note that the number density of the seed nuclei in the target is estimated by their atomic weight and the corresponding elemental mass density \cite{CRC97, RSC, ChemATE}.

%%%%%%=============================tables========================================

%%======section: target and data========
%%%%%%%%%%%%%%%%%%%%%%%%%%%%%%%%%%%%%%%%%%%%%%%%
\renewcommand{\arraystretch}{1.25}
\begin{longtable*}[!ht]{lcccccc}
%\begin{tabular}{lllllll}
\hline
\hline
Seed & $\lambda$ [mm] &   $T_{1/2}$ &   $T_{1/2}^{+1}$ &   $T_{1/2}^{+2}$ &   $T_{1/2}^{+3}$ &   $T_{1/2}^{+4}$ \tabularnewline
\hline
$^{7}_{3}\text{Li}$ &             32 &     stable &           840 ms &           178 ms &              - &           9 ms \tabularnewline
    \textbf{$^{9}_{4}\text{Be}$} &             14 &     stable &  \textbf{$2\times 10^6$ yr} &             14 s &          21 ms &     $2.7\times 10^{-15}$ ns \tabularnewline
             $^{11}_{5}\text{B}$ &             16 &     stable &            20 ms &            17 ms &          12 ms &          10 ms \tabularnewline
    \textbf{$^{13}_{6}\text{C}$} &             14 &     stable &  \textbf{$6\times 10^3$ yr} &              2 s &         747 ms &         193 ms \tabularnewline
             $^{15}_{7}\text{N}$ &          $5.2\times 10^4$ &     stable &              7 s &              4 s &         624 ms &         271 ms \tabularnewline
             $^{18}_{8}\text{O}$ &          $3.3\times 10^4$ &     stable &             27 s &             14 s &            3 s &            2 s \tabularnewline 
             $^{19}_{9}\text{F}$ &          $1.3\times 10^4$ &     stable &             11 s &              4 s &            4 s &            2 s \tabularnewline
           $^{22}_{10}\text{Ne}$ &          $7.8\times 10^4$ &     stable &             37 s &            3 min &         602 ms &         197 ms \tabularnewline
  \textbf{$^{23}_{11}\text{Na}$} &             51 &     stable &    \textbf{15 h} &             59 s &            1 s &         301 ms \tabularnewline
  \textbf{$^{26}_{12}\text{Mg}$} &             28 &     stable &            9 min &    \textbf{21 h} &            1 s &         335 ms \tabularnewline
           $^{27}_{13}\text{Al}$ &             16 &     stable &            2 min &            7 min &            4 s &         644 ms \tabularnewline
           $^{32}_{14}\text{Si}$ &             38 &      153 yr &              6 s &              3 s &         780 ms &         450 ms \tabularnewline
   \textbf{$^{31}_{15}\text{P}$} &             61 &     stable &    \textbf{14 d} &    \textbf{25 d} &           12 s &           47 s \tabularnewline
   \textbf{$^{36}_{16}\text{S}$} &             90 &     stable &            5 min &     \textbf{3 h} &           12 s &            9 s \tabularnewline
           $^{37}_{17}\text{Cl}$ &          $6.1\times 10^4$ &     stable &           37 min &           56 min &          1 min &           38 s \tabularnewline
  \textbf{$^{41}_{18}\text{Ar}$} &          $8.5\times 10^4$ &        2 h &    \textbf{33 yr} &            5 min &         12 min &           21 s \tabularnewline
   \textbf{$^{41}_{19}\text{K}$} &          $1.4\times 10^2$ &     stable &    \textbf{12 h} &    \textbf{22 h} &         22 min &         18 min \tabularnewline
           $^{48}_{20}\text{Ca}$ &          $1.3\times 10^2$ &     $2\times 10^{19}$ yr &            9 min &             14 s &           10 s &            5 s \tabularnewline
  \textbf{$^{45}_{21}\text{Sc}$} &             20 &     stable &    \textbf{84 d} &     \textbf{3 d} &   \textbf{2 d} &         57 min \tabularnewline
           $^{50}_{22}\text{Ti}$ &             28 &     stable &            6 min &            2 min &           33 s &            2 s \tabularnewline
            $^{51}_{23}\text{V}$ &             11 &     stable &            4 min &            2 min &           50 s &            7 s \tabularnewline
           $^{54}_{24}\text{Cr}$ &             19 &     stable &            3 min &            6 min &           21 s &            7 s \tabularnewline
  \textbf{$^{55}_{25}\text{Mn}$} &             15 &     stable &     \textbf{3 h} &            1 min &            3 s &            5 s \tabularnewline
  \textbf{$^{58}_{26}\text{Fe}$} &              5 &     stable &    \textbf{44 d} &  \textbf{$2\times 10^6$ yr} &          6 min &          1 min \tabularnewline
  \textbf{$^{59}_{27}\text{Co}$} &             13 &     stable &     \textbf{5 yr} &     \textbf{2 h} &          2 min &           27 s \tabularnewline
  \textbf{$^{64}_{28}\text{Ni}$} &              5 &     stable &     \textbf{3 h} &     \textbf{2 d} &           21 s &           29 s \tabularnewline
  \textbf{$^{65}_{29}\text{Cu}$} &             14 &     stable &            5 min &     \textbf{3 d} &           31 s &          3 min \tabularnewline
  \textbf{$^{70}_{30}\text{Zn}$} &             24 &  $>$$10^{16}$ yr &            2 min &     \textbf{2 d} &           24 s &          2 min \tabularnewline
  \textbf{$^{71}_{31}\text{Ga}$} &             26 &     stable &    \textbf{14 h} &     \textbf{5 h} &          8 min &          2 min \tabularnewline
  \textbf{$^{76}_{32}\text{Ge}$} &             30 &     stable &    \textbf{11 h} &     \textbf{1 h} &           19 s &           30 s \tabularnewline
  \textbf{$^{75}_{33}\text{As}$} &             23 &     stable &     \textbf{1 d} &     \textbf{2 d} &   \textbf{2 h} &          9 min \tabularnewline
           $^{82}_{34}\text{Se}$ &             32 &     stable &           22 min &            3 min &           32 s &           14 s \tabularnewline
  \textbf{$^{81}_{35}\text{Br}$} &             50 &     stable &     \textbf{1 d} &     \textbf{2 h} &         32 min &          3 min \tabularnewline
  \textbf{$^{86}_{36}\text{Kr}$} &          $2.4\times 10^4$ &     stable &     \textbf{1 h} &     \textbf{3 h} &          3 min &           32 s \tabularnewline
           $^{87}_{37}\text{Rb}$ &          $1.1\times 10^2$ &     $5\times 10^{10}$ yr &           18 min &           15 min &          3 min &           58 s \tabularnewline
  \textbf{$^{90}_{38}\text{Sr}$} &             60 &       29 yr &    \textbf{10 h} &     \textbf{3 h} &          7 min &          1 min \tabularnewline
   \textbf{$^{91}_{39}\text{Y}$} &             36 &       59 d &     \textbf{4 h} &    \textbf{10 h} &         19 min &         10 min \tabularnewline
  \textbf{$^{96}_{40}\text{Zr}$} &             21 &     $2\times 10^{19}$ yr &    \textbf{17 h} &             31 s &            2 s &            7 s \tabularnewline
  \textbf{$^{95}_{41}\text{Nb}$} &             20 &       35 d &    \textbf{23 h} &     \textbf{1 h} &            3 s &           15 s \tabularnewline
          $^{100}_{42}\text{Mo}$ &             19 &     $7\times 10^{18}$ yr &           15 min &           11 min &          1 min &          1 min \tabularnewline
           $^{99}_{43}\text{Tc}$ &             14 &     $2\times 10^5$ yr &             15 s &           14 min &            5 s &           54 s \tabularnewline
          $^{106}_{44}\text{Ru}$ &             19 &        1 yr &            4 min &            5 min &           34 s &           12 s \tabularnewline
          $^{105}_{45}\text{Rh}$ &             18 &        1 d &             30 s &           22 min &           17 s &          1 min \tabularnewline
 \textbf{$^{110}_{46}\text{Pd}$} &             20 &     stable &           23 min &    \textbf{21 h} &          2 min &          2 min \tabularnewline
          $^{113}_{47}\text{Ag}$ &             26 &        5 h &              5 s &           20 min &          4 min &          1 min \tabularnewline
 \textbf{$^{116}_{48}\text{Cd}$} &             28 &     $3\times 10^{19}$ yr &     \textbf{2 h} &           50 min &          3 min &           51 s \tabularnewline
          $^{115}_{49}\text{In}$ &             40 &     $4\times 10^{14}$ yr &             14 s &           43 min &            5 s &          2 min \tabularnewline
 \textbf{$^{126}_{50}\text{Sn}$} &             44 &     $2\times 10^5$ yr &     \textbf{2 h} &           59 min &          2 min &          4 min \tabularnewline
    \textbf{$^{126}_{51}\text{Sb}$} &             48 &       12 d &   \textbf{4 d} &    \textbf{9 h} &   \textbf{4 h} &         40 min \tabularnewline
              $^{132}_{52}\text{Te}$ &             38 &        3 d &         12 min &          42 min &           19 s &           18 s \tabularnewline
              $^{135}_{53}\text{I}$ &             64 &        7 h &          1 min &            24 s &            6 s &            2 s \tabularnewline
              $^{136}_{54}\text{Xe}$ &          $4.5\times 10^4$ &  $>$$2\times 10^{21}$ yr &          4 min &          14 min &           40 s &           14 s \tabularnewline
              $^{137}_{55}\text{Cs}$ &          $1.8\times 10^2$ &       30 yr &         33 min &           9 min &          1 min &           25 s \tabularnewline
              $^{140}_{56}\text{Ba}$ &             91 &       13 d &         18 min &          11 min &           14 s &           12 s \tabularnewline
    \textbf{$^{140}_{57}\text{La}$} &             54 &        2 d &   \textbf{4 h} &    \textbf{2 h} &         14 min &           41 s \tabularnewline
             $^{144}_{58}\text{Ce}$ &             45 &      285 d &          3 min &          14 min &           56 s &           56 s \tabularnewline
    \textbf{$^{143}_{59}\text{Pr}$} &             48 &       14 d &         17 min &    \textbf{6 h} &         24 min &         13 min \tabularnewline
             $^{150}_{60}\text{Nd}$ &             32 &     $8\times 10^{18}$ yr &         12 min &          11 min &           32 s &           26 s \tabularnewline
             $^{151}_{61}\text{Pm}$ &             32 &        1 d &          4 min &           5 min &          2 min &           42 s \tabularnewline
    \textbf{$^{154}_{62}\text{Sm}$} &             31 &     stable &         22 min &    \textbf{9 h} &          8 min &          5 min \tabularnewline
             $^{157}_{63}\text{Eu}$ &             42 &       15 h &         46 min &          18 min &           38 s &           26 s \tabularnewline
             $^{160}_{64}\text{Gd}$ &             35 &  $>$$3\times 10^{19}$ yr &          4 min &           8 min &          1 min &           45 s \tabularnewline
             $^{161}_{65}\text{Tb}$ &             28 &        7 d &          8 min &          20 min &          3 min &          2 min \tabularnewline
    \textbf{$^{164}_{66}\text{Dy}$} &             31 &     stable &   \textbf{2 h} &    \textbf{3 d} &          6 min &          9 min \tabularnewline
  \textbf{$^{166m1}_{67}\text{Ho}$} &             29 &     $1\times 10^3$ yr &   \textbf{3 h} &           3 min &          5 min &          3 min \tabularnewline
    \textbf{$^{170}_{68}\text{Er}$} &             32 &     stable &   \textbf{8 h} &    \textbf{2 d} &          1 min &          3 min \tabularnewline
    \textbf{$^{171}_{69}\text{Tm}$} &             30 &        2 yr &   \textbf{3 d} &    \textbf{8 h} &          5 min &         15 min \tabularnewline
    \textbf{$^{176}_{70}\text{Yb}$} &             44 &     stable &   \textbf{2 h} &    \textbf{1 h} &          8 min &          2 min \tabularnewline
    \textbf{$^{176}_{71}\text{Lu}$} &             28 &     $4\times 10^{10}$ yr &   \textbf{7 d} &          28 min &   \textbf{5 h} &          6 min \tabularnewline
    \textbf{$^{182}_{72}\text{Hf}$} &             26 &     $9\times 10^6$ yr &   \textbf{1 h} &    \textbf{4 h} &          4 min &          3 min \tabularnewline
    \textbf{$^{182}_{73}\text{Ta}$} &             18 &      115 d &   \textbf{5 d} &    \textbf{9 h} &         49 min &         10 min \tabularnewline
     \textbf{$^{186}_{74}\text{W}$} &             13 &  $>$$2\times 10^{20}$ yr &   \textbf{1 d} &   \textbf{70 d} &         11 min &         30 min \tabularnewline
    \textbf{$^{187}_{75}\text{Re}$} &              9 &     $4\times 10^{10}$ yr &  \textbf{17 h} &    \textbf{1 d} &          3 min &         10 min \tabularnewline
    \textbf{$^{192}_{76}\text{Os}$} &              9 &     stable &   \textbf{1 d} &    \textbf{6 yr} &          9 min &         35 min \tabularnewline
    \textbf{$^{193}_{77}\text{Ir}$} &             11 &     stable &  \textbf{19 h} &    \textbf{2 h} &           52 s &          6 min \tabularnewline
    \textbf{$^{198}_{78}\text{Pt}$} &             11 &     stable &         31 min &   \textbf{13 h} &          2 min &   \textbf{2 d} \tabularnewline
    \textbf{$^{197}_{79}\text{Au}$} &             14 &     stable &   \textbf{3 d} &    \textbf{3 d} &         48 min &         26 min \tabularnewline
             $^{204}_{80}\text{Hg}$ &             19 &     stable &          5 min &           8 min &          3 min &         41 min \tabularnewline
             $^{205}_{81}\text{Tl}$ &             20 &     stable &          4 min &           5 min &          3 min &          2 min \tabularnewline
    \textbf{$^{208}_{82}\text{Pb}$} &             20 &     stable &   \textbf{3 h} &   \textbf{22 yr} &         36 min &  \textbf{11 h} \tabularnewline
  \textbf{$^{210m1}_{83}\text{Bi}$} &             28 &     $3\times 10^6$ yr &          2 min &    \textbf{1 h} &         46 min &         20 min \tabularnewline
             $^{210}_{84}\text{Po}$ &             47 &      138 d &         516 ms &         0.30 ns &           4 ns &         164 ns \tabularnewline
    \textbf{$^{226}_{88}\text{Ra}$} &             54 &     $2\times 10^3$ yr &         42 min &    \textbf{6 yr} &          4 min &   \textbf{2 h} \tabularnewline
    \textbf{$^{227}_{89}\text{Ac}$} &             27 &       22 yr &   \textbf{6 h} &    \textbf{1 h} &          2 min &          8 min \tabularnewline
             $^{234}_{90}\text{Th}$ &             25 &       24 d &          7 min &          37 min &          5 min &          9 min \tabularnewline
    \textbf{$^{233}_{91}\text{Pa}$} &             19 &       27 d &   \textbf{7 h} &          24 min &          9 min &          9 min \tabularnewline
              $^{240}_{92}\text{U}$ &             16 &       14 h &          5 min &          17 min &              - &              - \tabularnewline
    \textbf{$^{239}_{93}\text{Np}$} &             15 &        2 d &   \textbf{1 h} &          14 min &          2 min &          2 min \tabularnewline
    \textbf{$^{246}_{94}\text{Pu}$} &             16 &       11 d &   \textbf{2 d} &               - &              - &              - \tabularnewline
    \textbf{$^{244}_{95}\text{Am}$} &             25 &       10 h &   \textbf{2 h} &          39 min &         23 min &         10 min \tabularnewline
             $^{250}_{96}\text{Cm}$ &             23 &     $8\times 10^3$ yr &         17 min &               - &              - &              - \tabularnewline
             $^{250}_{97}\text{Bk}$ &             21 &        3 h &         56 min &               - &         10 min &          2 min \tabularnewline
    \textbf{$^{254}_{98}\text{Cf}$} &             20 &       60 d &   \textbf{1 h} &          12 min &              - &              - \tabularnewline
    \textbf{$^{255}_{99}\text{Es}$} &             37 &       40 d &         25 min &    \textbf{8 d} &          2 min &              - \tabularnewline
   \textbf{$^{255}_{100}\text{Fm}$} &             37 &       20 h &   \textbf{3 h} &  \textbf{100 d} &         370 ns &            2 s \tabularnewline
\hline
\hline
%\label{tab:info}
%\end{tabular}
\caption{Seed nuclides considered in this study. The half life of the seed nuclides $T_{1/2}$ and of the corresponding $i$-species neutron-enriched isotopes $T_{1/2}^{+i}$ presented are taken from the ENDF-B-VIII.0 decay sublibrary \cite{ENDFBVIII0}. For the cases in which only a bound for the half life is available, this bound is used during all calculations, and similarly unknown half lives are treated as zero. The neutron penetration depth $\lambda$ is the minimal penetration depth in the energy (average neutron energy) range between $50$ keV and $10$ MeV for the seed material. Half lives fulfilling $T_{1/2}^{+i} > 1$ h are marked bold and correspondingly seed nuclides that have at least one such sufficiently-stable neutron-enriched isotope are also marked bold. }
\label{tab:info}
\end{longtable*}
\renewcommand{\arraystretch}{1}
%%%%%%%%%%%%%%%%%%%%%%%%%%%%%%%%%%%%%%%%%%%%%%%%
%%%%%%=============================tables========================================

\subsection{ENDF/B-VIII.0 data and NON-SMOKER data}

The ENDF/B-VIII.0 neutron sublibrary (denoted as ENDF-B) data comes in one data file (in the ENDF6 format \cite{ENDF6}) per nuclide containing a variety of different cross sections and other information, such as energies of excited states. Before usage the data needs to be preprocessed using the PREPRO \cite{PREPRO18} code to interpolate between the experimental data and add the contribution of resonances to the cross sections.

For the purpose of the present study the ENDF-B library provides us with cross sections for the considered seed nuclides. From the ENDF-B library, we obtain the total neutron interaction cross section $\sigma_{tot, 0}$ that includes all possible processes of elastic scattering ($n$, $n$), inelastic scatting with excitation of the nucleus ($n$, $n'$) and non-elastic scattering. Furthermore we employ the neutron-capture cross section $\sigma_{c, 0}$, the elastic cross section $\sigma_{el, 0}$, the inelastic cross section $\sigma_{inel, 0}$ and the non-elastic cross section $\sigma_{non-el, 0}$. This allows us to calculate the transmutation cross section $\sigma_{tr, 0}$, describing all neutron interaction processes changing the proton or neutron number of the target seed nuclei
\begin{equation}
  \sigma_{tr, 0} = \sigma_{non-el, 0} - \sigma_{inel, 0}.
\end{equation}

As the non-elastic cross section $\sigma_{non-el,0}$ is not available for most of the seed nuclides we study, we calculate it for the missing cases by means of the total and elastic cross section via
\begin{equation} \label{eq:nonelcal}
  \sigma_{non-el, 0} = \sigma_{tot, 0} - \sigma_{el, 0}.
\end{equation}
In order to check the precision of $\sigma_{non-el, 0}$ obtained by this procedure we compare it with the values directly taken from the ENDF-B library for the cases where this data exist. The comparison shows that the mean relative error of $\sigma_{non-el, 0}$ calculated by Eq.~(\ref{eq:nonelcal}) is around $1 \%$ for most cases. For a few cases, however, one finds a mean relative error of about $50 \%$. The error in the calculated non-elastic cross section is possibly due to miss alignments of the energy grids on which the different cross sections are given. Thus, we conclude that $\sigma_{non-el, 0}$ provided by Eq.~(\ref{eq:nonelcal}) is a good approximation.

Furthermore, we note that in the ENDF-B library, inelastic cross sections $\sigma_{inel,0}$ are missing for $^{18}$O, $^{13}$C, $^{233}$Pa and $^{9}$Be. However, for $^{18}$O, $^{13}$C and $^{233}$Pa inelastic cross section data is added during preprocessing by PREPRO.

As we select the potential seed nuclides by the heaviest sufficiently-long-lived isotope per element whose data is available in the ENDF-B library, the cross sections for almost all of the neutron-enriched isotopes are not available. We therefore also make use of theoretically predicted neutron-capture cross sections by the NON-SMOKER code \cite{RAUSCHER200147} for the neutron-capture cross sections $\sigma_{c,i}$ ($i>0$) of the neutron-enriched isotopes. While neutron cross sections are only provided for $10\leq Z \leq 83$, the range of covered neutron numbers in the NON-SMOKER data exceeds that of the ENDF-B data. The NON-SMOKER code predicts two sets of cross sections based on the Extended Thomas-Fermi Approach with Strutinski Integral model (ETFSI-Q) and the Finite Range Droplet Model (FRDM) respectively, where the first one only covers $26\leq Z \leq 83$.

In the NON-SMOKER library, the nuclear cross sections were calculated in the Hauser-Feshbach statistical model for compound nucleus reactions \cite{RAUSCHER200147, RauscherADNDT2000}. Two sets of cross sections are provided by the calculations with input from two different mass models: the FRDM and the ETFSI-Q. Experimental masses and level properties were used in the calculations where available. The cross sections in the NON-SMOKER library are of purely theoretical nature, as no direct experimental information was used in the calculations except for nuclear masses and ground and excited state information where available. The main goal of the NON-SMOKER library is to provide consistent cross sections for all nuclides from the valley of stability to the neutron and proton drip lines, which are astrophysical relevant. As the Hauser-Feshbach statistical model for nuclear reactions is used, the NON-SMOKER library is applicable for nuclear reactions with the compound nucleus reaction mechanism \cite{RAUSCHER200147, RauscherADNDT2000}. This is the case for high-level densities with completely overlapping resonances \cite{RAUSCHER200147, RauscherADNDT2000}. In general, the compound nucleus mechanism only dominates for relatively low energies smaller than $20$ MeV \cite{RAUSCHER200147, RauscherADNDT2000}. This energy condition is fulfilled for our calculations as we consider the energy range of $50$ keV to $10$ MeV for incident neutrons. However, for light nuclei or decreasing particle separation energies or at shell closures, level densities would become too low for the application of the statistical model at the energies under consideration \cite{RAUSCHER200147, RauscherADNDT2000}. However, even with those inaccuracies and weaknesses, the cross sections calculated with NON- SMOKER deviate from experimental data by a factor of about $1.3$-$1.4$ on average for neutron capture cross sections\cite{RAUSCHER200147, RauscherADNDT2000}. Furthermore, for the cases of neutron-rich nuclides we are interested in, comparison with the predictions of cross sections from TALYS-based evaluated nuclear data library (TENDL) \cite{KoningNDS2012, TENDLweb2017, TENDLweb2019} where other nuclear models are included, shows that dramatical changes are not expected.

Comparisons of the neutron capture cross sections from the ENDF-B library and the NON-SMOKER data show that the latter mostly fit the data from ENDF-B within one order of magnitude, except at resonances, where they just follow the value averaged over the resonance. For a quantitative comparison we look at the $N_1$ abundances obtained from the neutron-capture cross sections taken from the NON-SMOKER data and the ENDF-B data, respectively. The relative deviations of the two results for $N_1$ after one pulse are computed for $300$ log-spaced energies (average incident neutron energy) between $50$ keV to $10$ MeV and for different seed nuclides. Both models of the NON-SMOKER data perform equally well and have a mean relative deviation to the ENDF-B result of around $50 \%$. The details of the comparison are shown in the Appendix~\ref{sec:apddata}. In the present work, the FRDM data is employed for the neutron-capture cross sections $\sigma_{c,i}$ with $i>0$, as it covers a larger range of elements than the ETFSI-Q data. We note that if no neutron-capture cross section is available in the NON-SMOKER data, the last available cross section (less neutrons) is used as an approximation, $\mu_i = \mu_{i-1}$. For the elements with $Z<10$ or $Z>83$ this leads to the approximation $\mu_4 \approx \mu_3 \approx \mu_2 \approx \mu_1 \approx \mu_0$.

The non-elastic and inelastic cross sections of neutron-enriched isotopes are estimated by the one of the seed nuclide as they are not available for almost all neutron-enriched isotopes in the ENDF-B or NON-SMOKER data (only for the seed nuclides $^{113}_{47}$Ag, $^{193}_{77}$Ir and $^{240}_{92}$U, non-elastic and inelastic cross section data is available for some of their neutron-enriched isotopes in the ENDF-B). Analysis of the non-elastic cross sections for the seed nuclide shows that, up to a few $100$ keV the non-elastic cross section is dominated by neutron capture for most (non fissile) nuclides. Above this energy scale, excitations of the nucleus and then high energy processes such as proton production, etc. become dominating. We therefore estimate for $i\geq 1$
\begin{equation} \label{eq:etapprox}
\eta_i = \mu_i + \eta_0 - \mu_0.  
\end{equation}
For the situation that the neutron capture might no be the dominating process of the non-elastic cross section, the above approximation (\ref{eq:etapprox}) is questionable, especially for fissile isotopes, such as some isotopes of the Actinides. However, comparison with the predictions of cross sections from nuclear model calculations, such as the TALYS-based evaluated nuclear data library (TENDL) \cite{KoningNDS2012, TENDLweb2017, TENDLweb2019}, shows that dramatical changes are not expected for the cases we are interested in. However, we note that cross sections predicted from nuclear model calculations, such as TENDL \cite{KoningNDS2012, TENDLweb2017, TENDLweb2019} and the FRESCO code \cite{ThompsonCPR1998, FrescoWeb2020}, may help to improve the precision of the calculation.

\subsection{Penetration depths}

In order to have an idea of the damping of the incident neutron beam inside the target, we present the minimal penetration depth for all potential seed materials in Table~(\ref{tab:info}). By means of Eq.~(\ref{eq:pend}) the penetration depth is calculated over an average incident neutron energy range of $50$ keV to $10$ MeV for each seed nuclide. Afterwards the minimal value per seed nuclide is selected for presentation, since the smallest penetration depths have the strongest effect on the actual quantities ($N_i$ and $N^\text{tot}_i$) we are interested in. As the minimal penetration depth found is about $5$ mm, the effect of the thickness of the target on the abundances is small ($\sim 1\%$ or less).

We note that the model in the present work is a 1D model without the detailed neutron transport. There are two types of 3D effects which would affect our results. The first one is the detailed interaction of the neutron with the target which can be described by neutron transport, including the loss of the incident neutrons and the change of the energy-momentum and position distribution of the neutrons. In our 1D model, these effects are only considered by the damping of the incident neutron beam. As we are analysing a thin target, our calculation on the penetration depths shows that only a very small fraction ($\sim 1\%$ or less) of the incident neutrons interacts with the target. As our result strongly depends on the flux of the neutron beam, especially for neutron-enriched isotopes, the small fraction of the scattered neutrons only has minor effects on the final abundance of neutron-enriched isotopes. Thus, the detailed neutron transport in the target would only lead to a deviation less than $1\%$. Another 3D effect is related to the angle divergence of the neutron beam and the target configuration. However, this depends very much on the neutron beam and the target configuration itself. Assuming there is no angel divergence of the neutron beam, then our 1D model is fine. Assuming the neutron source is completely isotropic, then the 3D effect would lead to the result of about $1$ order of magnitude less for the production of neutron-rich isotopes, due the to decrease of the neutron flux along the target thickness.

%%%%%%=============================figures========================================

%%======section: result one pulse========

%%%%%%%%%%%%%%%%%%%%%%%%%%%%%%%%%%%
\begin{figure*}[!ht]
  \includegraphics[width=\linewidth]{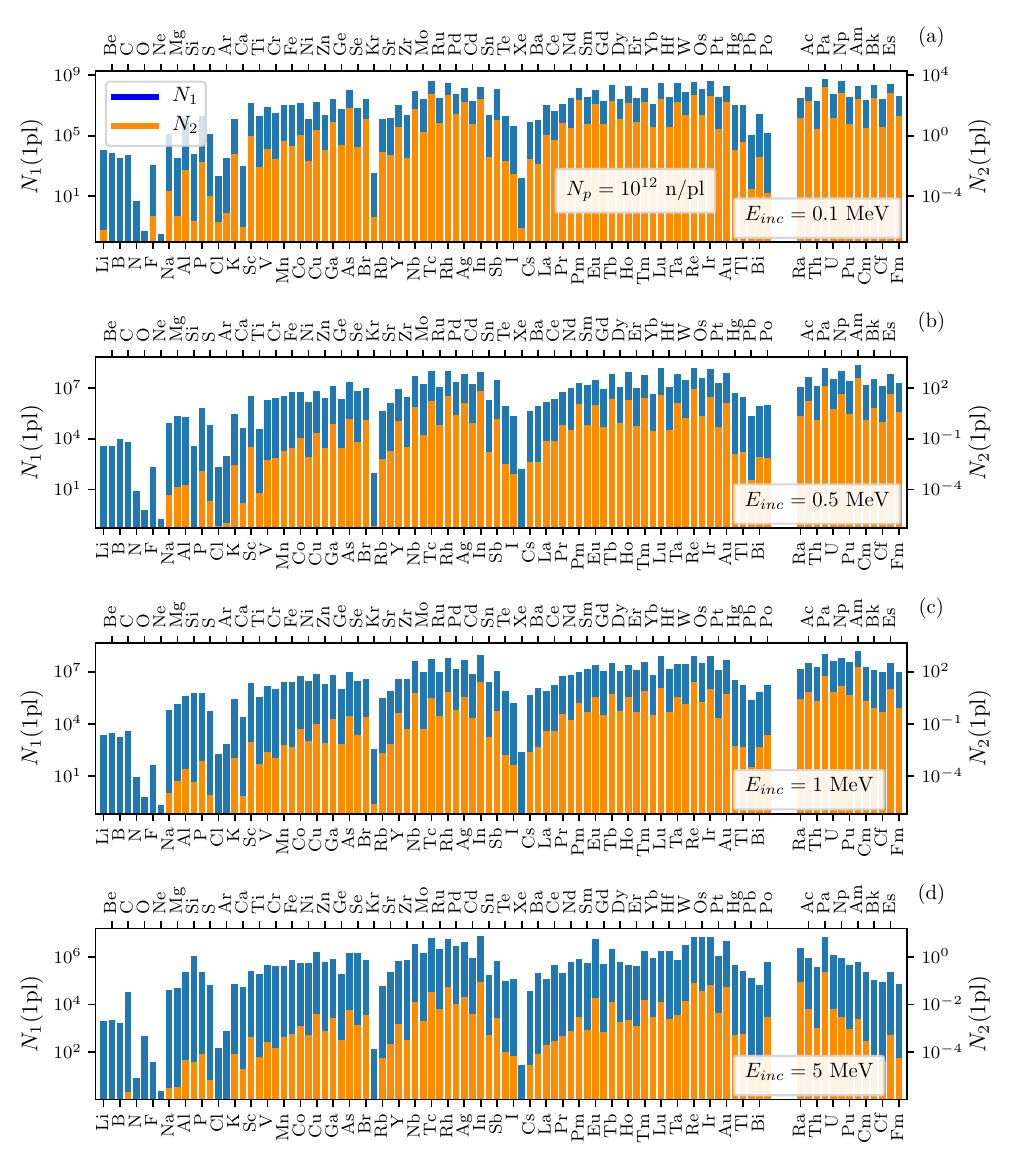}
  \caption{$N_1$ and $N_2$ after one neutron pulse, without taking the radioactive decay of the nuclei into account since $\tau_i \gg T_p$, for the seed nuclide listed in Table~\ref{tab:info} and selected average neutron energies: (a) $E_{inc}=100$ keV, (b) $E_{inc}=500$ keV, (c) $E_{inc}=1$ MeV and (d) $E_{inc}=5$ MeV. A neutron beam with $N_p=10^{12}$ n/pl is assumed. $N_1$ is represented by the blue bars, where their values need to be read from the left axis. $N_2$ is represented by the orange bars, where their values need to be read from the right axis. For some seed nuclides, $N_2$($1$pl) is not shown, because it is too small to be visible in the range shown in the plots.}
  \label{fig:1pl_N12}
\end{figure*}
\clearpage
%%%%%%%%%%%%%%%%%%%%%%%%%%%%%%%%%%%

%%%%%%%%%%%%%%%%%%%%%%%%%%%%%%%%%%%
\begin{figure*}[!ht]
  \includegraphics[width=\linewidth]{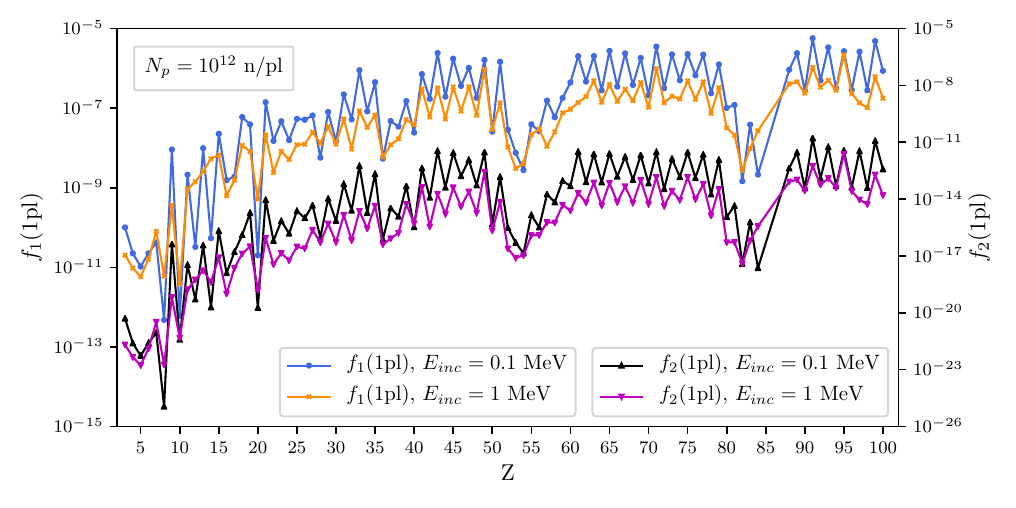}
  \caption{The fractions $f_1$ and $f_2$ of the number of $1$-species and $2$-species isotopes to the number of the target seed nuclei after one neutron pulse, without taking the radioactive decay of the nuclei into account since $\tau_i \gg T_p$, as functions of the atomic number $Z$ for the seed nuclide listed in Table~\ref{tab:info}. A neutron beam with $N_p=10^{12}$ n/pl is assumed and results for an average neutron energies $E_{inc}=100$ keV ($f_1$: blue curve with filled circles; $f_2$: black curve with filled up-pointing triangles) and $E_{inc}=1$ MeV ($f_1$: orange curve with crosses; $f_2$: purple curve with filled down-pointing triangles) are shown.}
  \label{fig:1pl_f12}
\end{figure*}
%%%%%%%%%%%%%%%%%%%%%%%%%%%%%%%%%%%

%%%%%%%%%%%%%%%%%%%%%%%%%%%%%%%%%%%
\begin{figure*}[!htb]
  \includegraphics[width=\linewidth]{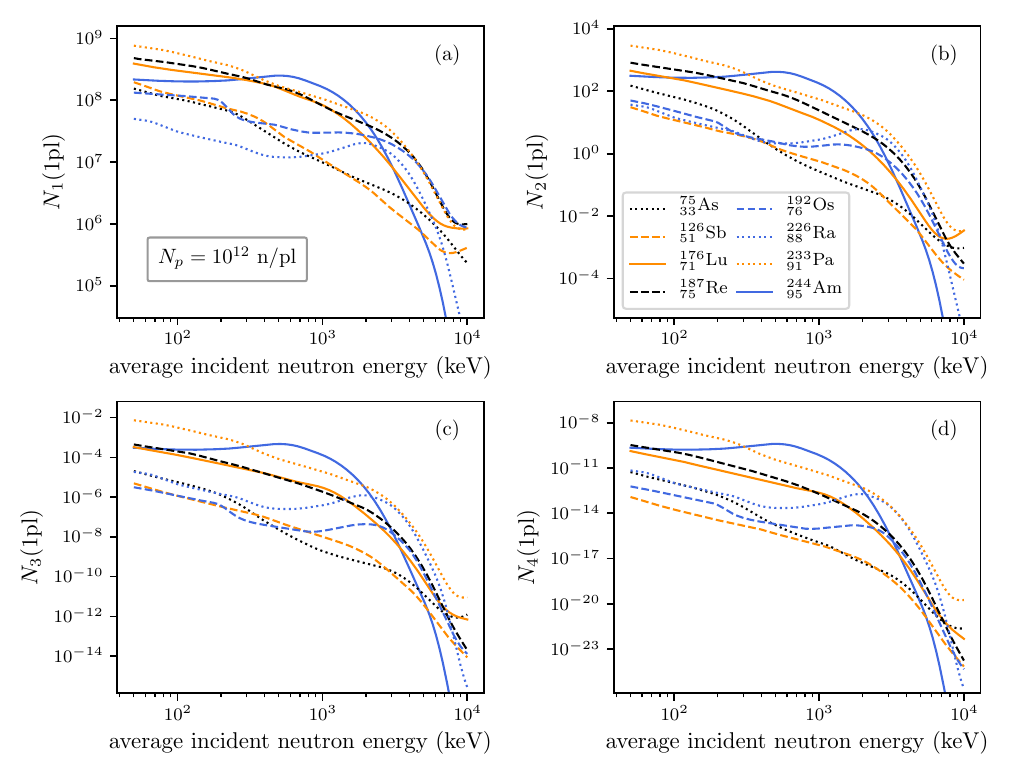}
  \caption{$N_1$ (a), $N_2$ (b), $N_3$ (c), and $N_4$ (d) after one neutron pulse, without taking the radioactive decay of the nuclei into account since $\tau_i \gg T_p$, as functions of the average incident neutron energy for a selection of seed nuclides. A neutron beam with $N_p=10^{12}$ n/pl is assumed. The seed nuclides $^{75}_{33}$As (black dotted curve), $^{126}_{51}$Sb (orange dashed curve), $^{176}_{71}$Lu (orange solid curve), $^{187}_{75}$Re (black dashed curve), $^{192}_{76}$Os (blue dashed curve), $^{226}_{88}$Ra (blue dotted curve), $^{233}_{91}$Pa (orange dotted curve) and $^{244}_{95}$Am (blue solid curve) are considered.}
  \label{fig:best_1pl}
\end{figure*}
%%%%%%%%%%%%%%%%%%%%%%%%%%%%%%%%%%%

%%%%%%=============================figures========================================

%%%%%%=============================figures========================================

%%======section: result 10000 pulse========

%%%%%%%%%%%%%%%%%%%%%%%%%%%%%%%%
\begin{figure*}[!ht]
  \includegraphics[width=\linewidth]{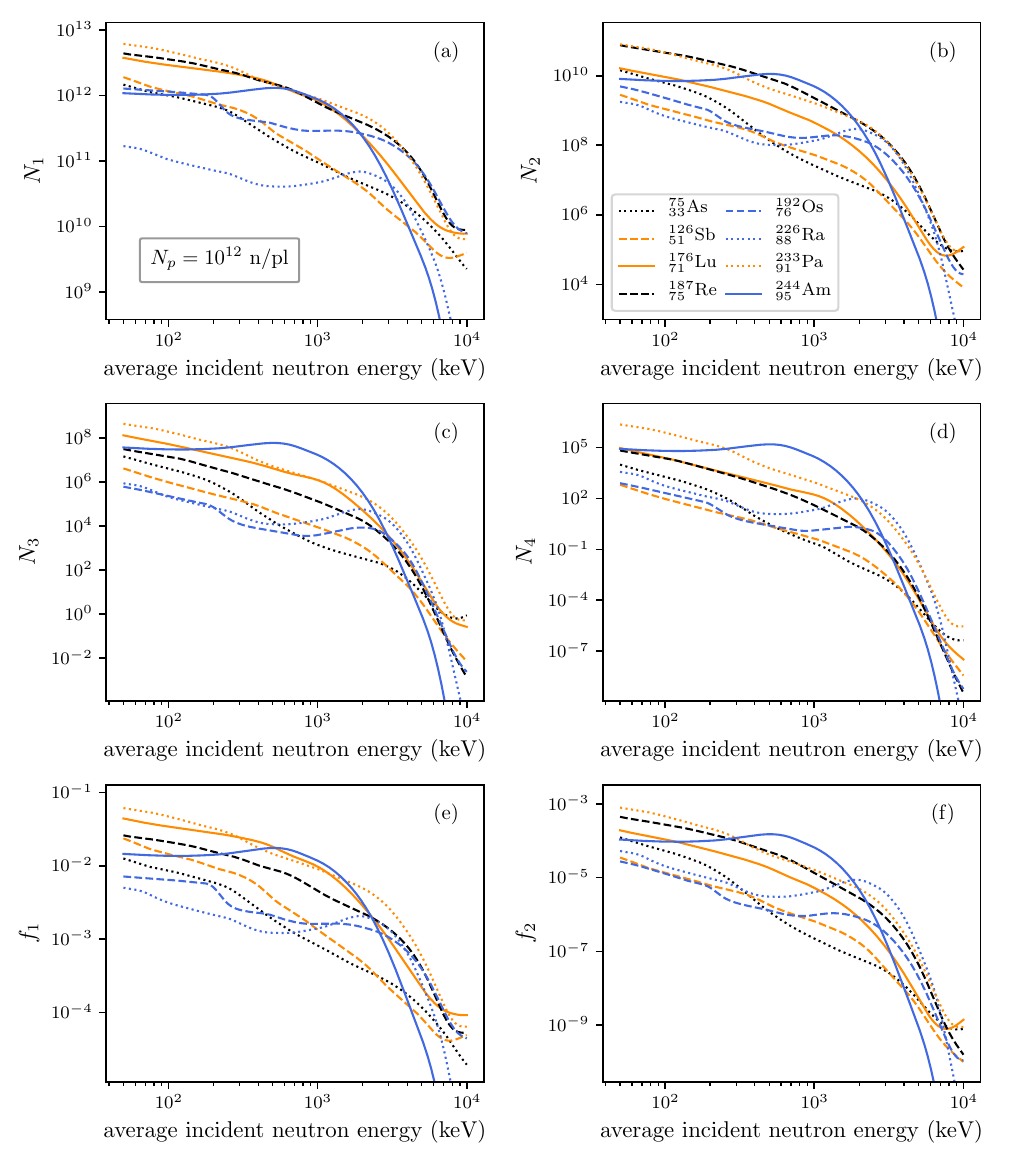}
  \caption{$N_i$ [(a), (b), (c), and (d)] and the fraction $f_i$ [(e) and (f)] after $10^4$ neutron pulses (at time $10^4 T_{del}$) as functions of the average incident neutron energy. A neutron beam with $N_p=10^{12}$ n/pl and a repetition rate of $f_{rep} = 1$ Hz is assumed. The seed nuclides $^{75}_{33}$As (black dotted curve), $^{126}_{51}$Sb (orange dashed curve), $^{176}_{71}$Lu (orange solid curve), $^{187}_{75}$Re (black dashed curve), $^{192}_{76}$Os (blue dashed curve), $^{226}_{88}$Ra (blue dotted curve), $^{233}_{91}$Pa (orange dotted curve) and $^{244}_{95}$Am (blue solid curve) are considered here.}
  \label{fig:best_abu_fN_res12}
\end{figure*}
%%%%%%%%%%%%%%%%%%%%%%%%%%%%%%%%

%%%%%%%%%%%%%%%%%%%%%%%%%%%%%%%%%%%%%
\begin{figure*}[!ht]
  \includegraphics[width=\linewidth]{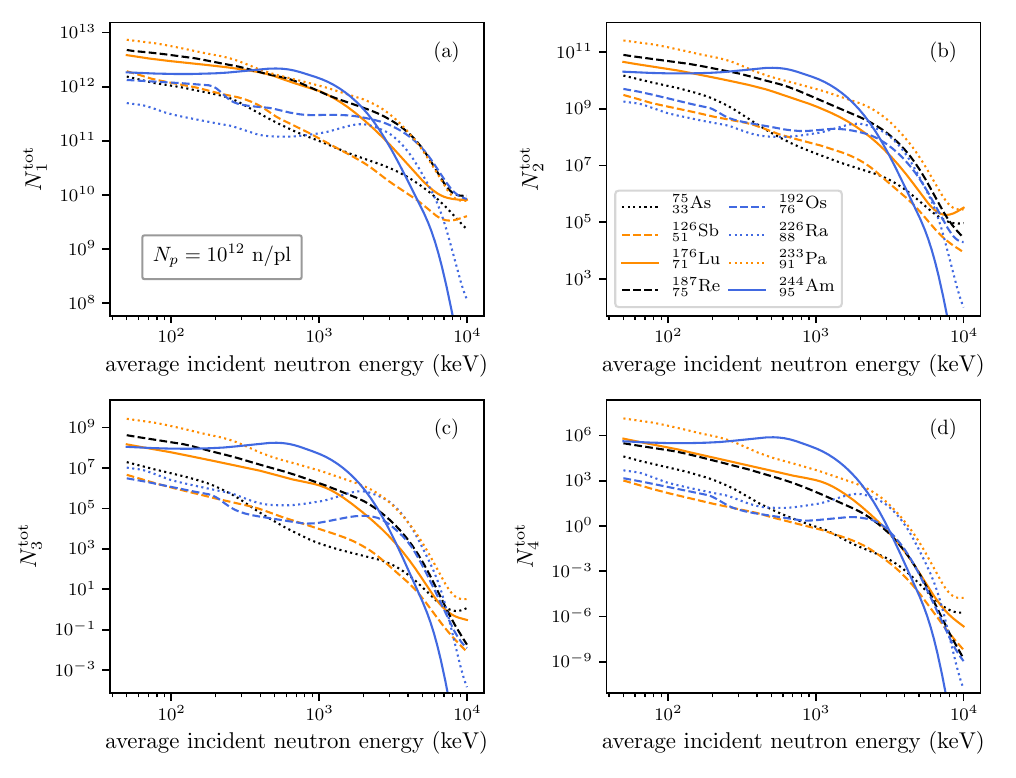}
  \caption{Total amount of the neutron-enriched nuclei $N_i^{\text{tot}}$ after $10^4$ neutron pulses as functions of the average incident neutron energy. A neutron beam with $N_p=10^{12}$ n/pl and a repetition rate of $f_{rep} = 1$ Hz is assumed. The seed nuclides $^{75}_{33}$As (black dotted curve), $^{126}_{51}$Sb (orange dashed curve), $^{176}_{71}$Lu (orange solid curve), $^{187}_{75}$Re (black dashed curve), $^{192}_{76}$Os (blue dashed curve), $^{226}_{88}$Ra (blue dotted curve), $^{233}_{91}$Pa (orange dotted curve) and $^{244}_{95}$Am (blue solid curve) are considered.}
  \label{fig:best_prod_N14_res12}
\end{figure*}
%%%%%%%%%%%%%%%%%%%%%%%%%%%%%%%%%%%%%

%%%%%%%%%%%%%%%%%%%%%%%%%%%%%%%%
\begin{figure*}[!ht]
  \includegraphics[width=\linewidth]{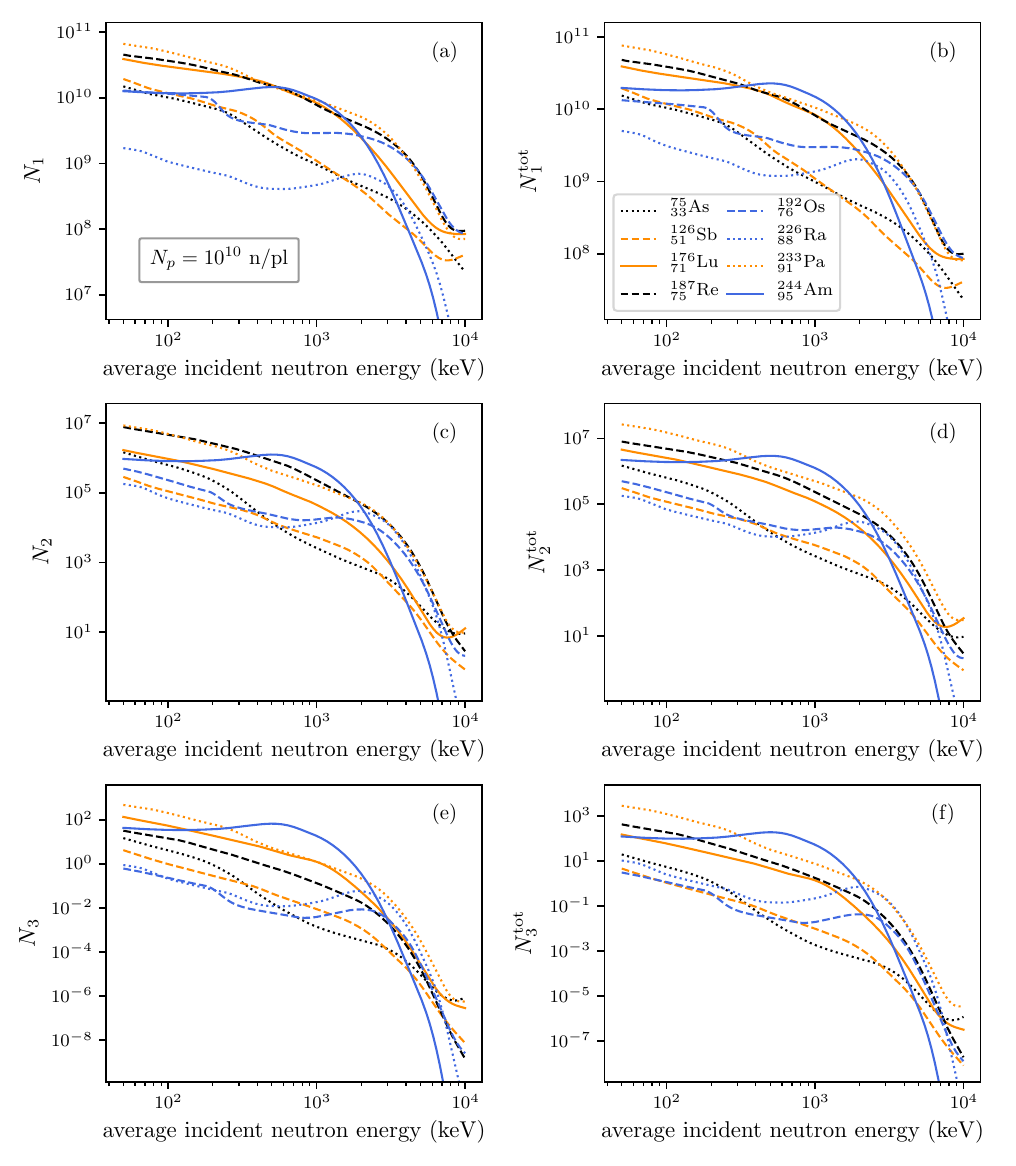}
  \caption{$N_i$ after $10^4$ neutron pulses (at time $10^4 T_{del}$) and total produced amount of neutron-enriched isotopes $N_i^{\text{tot}}$ as functions of the average incident neutron energy. A neutron beam with $N_p=10^{10}$ n/pl and a repetition rate of $f_{rep} = 1$ Hz is assumed. The seed nuclides $^{75}_{33}$As (black dotted curve), $^{126}_{51}$Sb (orange dashed curve), $^{176}_{71}$Lu (orange solid curve), $^{187}_{75}$Re (black dashed curve), $^{192}_{76}$Os (blue dashed curve), $^{226}_{88}$Ra (blue dotted curve), $^{233}_{91}$Pa (orange dotted curve) and $^{244}_{95}$Am (blue solid curve) are considered.}
  \label{fig:best_abu_prod_N_res10}
\end{figure*}
%%%%%%%%%%%%%%%%%%%%%%%%%%%%%%%%

%%%%%%%%%%%%%%%%%%%%%%%%%%%%%%%%%%%%
\begin{figure*}[!ht]
  \includegraphics[width=\linewidth]{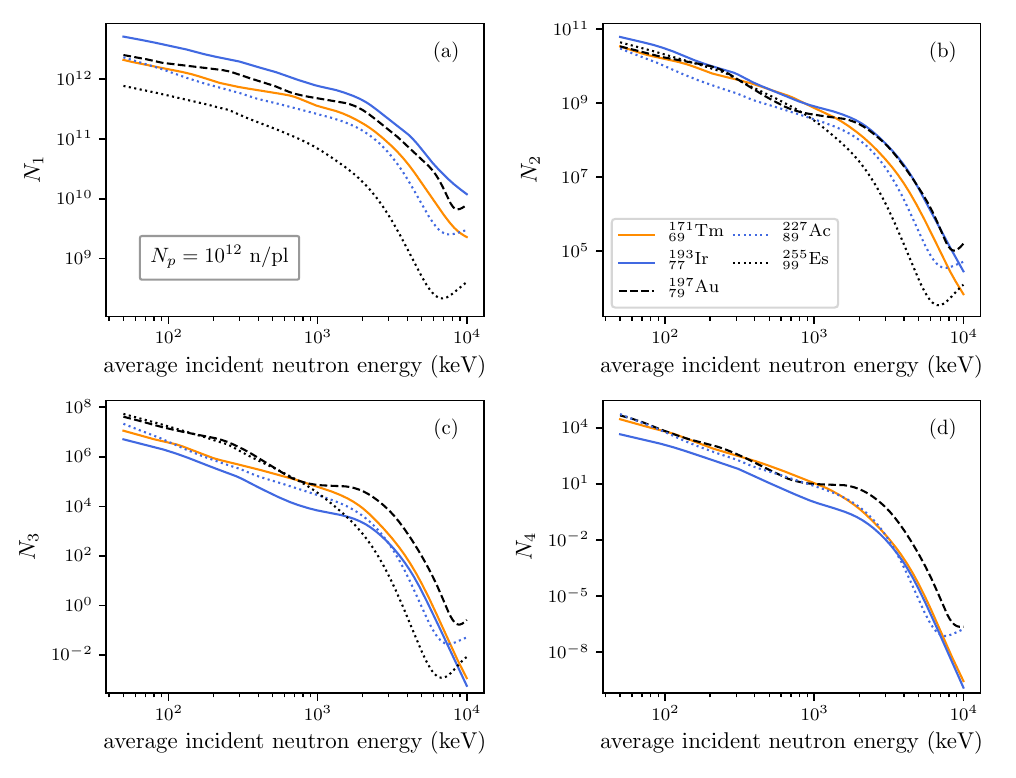}
  \caption{$N_i$ ($i$ up to $4$) after $10^4$ neutron pulses (at time $10^4 T_{del}$) as functions of the average incident neutron energy. A neutron beam with $N_p=10^{12}$ n/pl and a repetition rate of $f_{rep} = 1$ Hz is assumed. The seed nuclides $^{171}_{69}$Tm (orange solid curve), $^{193}_{77}$Ir (blue solid curve), $^{197}_{79}$Au (black dashed curve), $^{227}_{89}$Ac (blue dotted curve) and $^{255}_{99}$Es (black dotted curve) are considered.}
  \label{fig:sec_abu_prod_N_res12}
\end{figure*}
%%%%%%%%%%%%%%%%%%%%%%%%%%%%%%%%%%%%

%%%%%%=============================figures========================================

%%----------------------------------------------------------result one pulse-----------------------------------------------------------

\section{Numerical results for one neutron pulse \label{sec:nrone}}

In order to acquire an overall picture, we first consider the case of a single neutron pulse for each of the seed nuclides listed in Table~\ref{tab:info}. The results for $N_1$($1$pl) and $N_2$($1$pl) without accounting for any decay of the nuclei ($\tau_i \gg T_p$), calculated by Eq.~(\ref{eq:1pl_ex}), are presented in Fig.~\ref{fig:1pl_N12} for the average neutron energies $100$ keV, $500$ keV, $1$ MeV, and $5$ MeV. We assume a neutron beam with $N_p=10^{12}$ n/pl. The results show that the number of neutron capture events decreases in general as the incident neutron energy increases. This is due to the decreasing neutron-capture cross section. At an average neutron energy of $E_{inc}=100$ keV, $N_1$($1$pl) reaches an order of $10^9$ and $N_2$($1$pl) reaches an order of $10^3$, for some of the seed nuclides. For an average neutron energy of $E_{inc}=1$ MeV, $N_1$($1$pl) reaches an order of $10^8$, while $N_2$($1$pl) reaches an order of $10^2$. This indicates that even with one neutron pulse,  it is possible to observe $2$ successive neutron capture events.

Another important quantity is
\begin{equation}
  f_i = N_i/N_t,
\end{equation}
describing the fraction of the seed nuclei transmuting to the $i$-species neutron-enriched nuclei to the number of the seed nuclei $N_t$. The results for $f_1$($1$pl) and $f_2$($1$pl), after one neutron pulse and without taking the decay of nuclei into account ($\tau_i \gg T_p$), are shown in Fig.~\ref{fig:1pl_f12} as functions of the atomic number $Z$ of the seed nuclide. As in Fig.~\ref{fig:1pl_N12}, we assume again a neutron beam with $N_p=10^{12}$ n/pl. One observes in Fig.~\ref{fig:1pl_f12} that $f_1$($1$pl) and $f_2$($1$pl) have $3$ terrace regions in $Z$ where $f_i$($1$pl) have similar and quite good values: $41-51$, $60-79$, and $88-100$.

Among the good seed-candidates showing large $N_1$($1$pl) and $N_2$($1$pl) in Fig.~\ref{fig:1pl_N12} and Fig.~\ref{fig:1pl_f12}, seed nuclides which have at least one sufficiently-long-lived neutron-enriched isotope ($T_{1/2}^{+i} > 1$ h) are of particular interest. On the one hand, they provide the possibility of conducting further experimental studies of the nuclear properties of their neutron-enriched isotopes. On the other hand, they have advantages for the case of multiple neutron pulses, as the abundance of long-lived isotopes can accumulate in the target, thus enhancing the successive neutron capture, simultaneously allowing for the production of further neutron-enriched isotopes. In Fig.~\ref{fig:best_1pl}, we present the results of $N_1$($1$pl), $N_2$($1$pl), $N_3$($1$pl) and $N_4$($1$pl), again calculated by Eq.~(\ref{eq:1pl_ex}), as functions of the average neutron energy for a selection of such seed nuclides ($^{75}_{33}$As, $^{126}_{51}$Sb, $^{176}_{71}$Lu, $^{187}_{75}$Re, $^{192}_{76}$Os, $^{226}_{88}$Ra, $^{233}_{91}$Pa and $^{244}_{95}$Am). We once more assume a neutron beam with $N_p=10^{12}$ n/pl.

As shown in Fig.~\ref{fig:best_1pl}, $N_1$($1$pl) and $N_2$($1$pl) have values larger than $1$ for average incident neutron energies up to a few MeV. This, in principle, suggests the possibility of observing isotopes with $1$ and $2$ more neutrons than the seed nuclide after only a single neutron pulse. For all seeds and considered neutron energies, $N_3$($1$pl) and $N_4$($1$pl) are smaller than $1$, indicating that the successive neutron capture process of capturing more than $2$ neutrons is negligible in the single neutron pulse case. Figure~\ref{fig:best_1pl} also shows that the seed nuclides $^{233}_{91}$Pa, $^{187}_{75}$Re and $^{176}_{71}$Lu show in general a good performance in the considered neutron energy range ($50$ keV to $10$ MeV). Furthermore, $^{244}_{95}$Am performs well around a neutron energy of $1$ MeV. The seed nuclides $^{192}_{76}$Os and $^{226}_{88}$Ra exhibit a good performance for neutron energies of a few MeV, while they show a not as good performance at other neutron energies.

%%----------------------------------------------------------result 10000 pulses-----------------------------------------------------------

\section{Numerical results for multiple neutron pulses \label{sec:nrmul}}

We now turn to the scenario consisting of multiple neutron pulses, assuming a repetition rate of $f_{rep} = 1$ Hz for the neutron source, corresponding to $T_{del} = 1$ s. The target is exposed to $10^{4}$ neutron pulses ($N_{pl} = 10^4$), equivalent to roughly three hours of interaction time. As discussed previously, seed nuclides with at least one sufficiently-long-lived neutron-enriched isotope ($T_{1/2}^{+i} > 1$ h) are of primary interest and hence we restrict our study to such seed nuclides, marked bold in Table~\ref{tab:info}. 
We first consider a neutron beam with $N_p=10^{12}$ n/pl. The abundance $N_i$ and the total produced amount $N_i^{\text{tot}}$ after $10^4$ neutron pulses (at time $10^4 T_{del}$) is calculated by Eq.~(\ref{eq:nik}) and Eq.~(\ref{eq:ntik}), respectively. After analyses of the results we select a set of interesting seed nuclides ($^{75}_{33}$As, $^{126}_{51}$Sb, $^{176}_{71}$Lu, $^{187}_{75}$Re, $^{192}_{76}$Os, $^{226}_{88}$Ra, $^{233}_{91}$Pa and $^{244}_{95}$Am), which exhibit good performances. The results for $N_i$ ($i$ up to $4$) and the fractions $f_i$ ($i$ up to $2$) as functions of the average incident neutron energy are presented in Fig.~\ref{fig:best_abu_fN_res12}. Figure~\ref{fig:best_prod_N14_res12} shows $N_i^{\text{tot}}$ ($i$ up to $4$) in a similar manner. Again the seed nuclides $^{233}_{91}$Pa, $^{187}_{75}$Re and $^{176}_{71}$Lu exhibit a good performance for the entire neutron energy range we consider, $^{244}_{95}$Am performs well at neutron energies around $1$ MeV and $^{192}_{76}$Os and $^{226}_{88}$Ra show a good performance for neutron energies of a few MeV. We note that only the seed nuclides $^{75}_{33}$As, $^{126}_{51}$Sb and $^{176}_{71}$Lu have sufficiently-stable $3$-species isotopes ($T_{1/2}^{+3} > 1$ h), while $^{226}_{88}$Ra has  the $4$-species isotope with a half life of $T_{1/2}^{+4} > 1$ h.

After $10^4$ neutron pulses and at an average neutron energy of $\sim 100$ keV, approximatively $1\%$ to $10\%$ of the seed nuclei can be neutron-enriched, as shown in Fig.~\ref{fig:best_abu_fN_res12}. $N_3$ and $N_4$ remain observable ($>1$) up to a few MeV, indicating the possibility of producing isotopes with $4$ more neutrons than the seed nuclide. A comparison between Fig.~\ref{fig:best_abu_fN_res12} and Fig.~\ref{fig:best_prod_N14_res12} shows the effect of the decay and transmutation of nuclei, indicating that more neutron-enriched nuclei have been produced during the interaction than the remain at the end. As shown in Fig.~\ref{fig:best_prod_N14_res12}, $N_4^{\text{tot}}$ is still observable ($>1$) up to an average neutron energy of a few MeV. This, in principle, implies that it is possible to observe $4$ successive neutron capture events.

Comparisons of  the results in Fig.~\ref{fig:best_prod_N14_res12} with the results in Fig.~\ref{fig:best_1pl} show that, the production of the nuclei of the $i$-species isotope, for $i\geq 2$, depends on the number of neutron pulses $N_{pl}$ in a non-linear manner. The dependance of $N_{i}^{\text{tot}}$ is approximatively given by $N_{i}^{\text{tot}} \propto (N_{pl})^i$ [compare also to Eq.~(\ref{eq:N_scaling})]. This interesting feature shows the advantage of multiple neutron pulses, where such nonlinear increasing of $N_{i}^{\text{tot}}$ for $i>1$ is due to the accumulation of neutron-enriched nuclei during the multiple neutron pulses.

Furthermore, for all of these seed nuclides of the selected set ($^{75}_{33}$As, $^{126}_{51}$Sb, $^{176}_{71}$Lu, $^{187}_{75}$Re, $^{192}_{76}$Os, $^{226}_{88}$Ra, $^{233}_{91}$Pa and $^{244}_{95}$Am), the maximal $N_i$ ($i$ up to $4$) is always reached at the end of the $10^4$ neutron pulses. This indicates that even higher $N_i$ could be obtained by simply increasing the number of neutron pulses until saturation occurs when the contributions of neutron capture, radioactive decay and transmutation balance each other. The $(N_{pl})^i$ dependance of $N_{i}^{\text{tot}}$ remains until increasing the number of neutron pulses or increasing the parameter $N_p/A$ lead to the described saturation. Results in Fig.~\ref{fig:best_prod_N14_res12}(c), Fig.~\ref{fig:best_prod_N14_res12}(d) and Fig.~\ref{fig:best_1pl}(d) however indicate some sort of saturation for the production of nuclei of the $4$-species isotope for $^{187}_{75}$Re, $^{226}_{88}$Ra and $^{233}_{91}$Pa, as the scaling for $N_4^{\text{tot}}$ in these cases is slightly weaker than $(N_{pl})^i$. This is due to the quite short lifetime of the isotopes of the $3$-species of these seed nuclides.

In order to understand the effect of the number of neutrons per pulse $N_p$ and the parameter $N_p/A$, respectively, we also perform calculations for the case of $N_p = 10^{10}$ n/pl. The results for $N_i$ and $N_i^{\text{tot}}$ ($i$ up to $3$) after $10^4$ neutron pulses (at time $10^4 T_{del}$) as functions of the average neutron energy are shown in Fig.~\ref{fig:best_abu_prod_N_res10}, again for the same set of seed nuclides ($^{75}_{33}$As, $^{126}_{51}$Sb, $^{176}_{71}$Lu, $^{187}_{75}$Re, $^{192}_{76}$Os, $^{226}_{88}$Ra, $^{233}_{91}$Pa and $^{244}_{95}$Am). Both $N_i$ and $N_i^{\text{tot}}$ decrease compared to the $N_p = 10^{12}$ n/pl results and consequently have observable values for $N_3$ and $N_3^{\text{tot}}$ only up to a few $100$ keV of the average neutron energy. One finds that $N_i$ and $N_i^{\text{tot}}$ scale as $N_p^i$, since the parameter $N_p/A$ and the number of the neutron pulses $N_{pl}$ are too small to lead to saturation. This scaling can also be anticipated from the qualitative discussion leading to Eq.~(\ref{eq:N_scaling}). As discussed in Sec.~\ref{sec:seednuc}, $N_i/A$ is a function of $N_p/A$. Thus,  $N_i$ and $N_i^{\text{tot}}$ scale as $(N_p)^i/A^{i-1}$.

As mentioned above, we analyze all seed nuclides marked bold in Table~\ref{tab:info} having at least one sufficiently-long-lived neutron-enriched isotope ($T_{1/2}^{+i} > 1$ h). Beside the set of seed nuclides discussed so far and presented in Figs.~\ref{fig:best_abu_fN_res12}-\ref{fig:best_abu_prod_N_res10}, there are also some other seed nuclides exhibiting good results. We therefore also present results for the seed nuclides $^{171}_{69}$Tm, $^{193}_{77}$Ir, $^{197}_{79}$Au, $^{227}_{89}$Ac and $^{255}_{99}$Es. The corresponding values for $N_i$ ($i$ up to $4$) after $10^4$ neutron pulses (at time $10^4 T_{del}$) are shown in Fig.~\ref{fig:sec_abu_prod_N_res12} as functions of the average incident neutron energy. Here, we again assume a neutron beam with $N_p=10^{12}$ n/pl. While these nuclides perform worse than the best ones shown in Figs.~\ref{fig:best_abu_fN_res12}-\ref{fig:best_prod_N14_res12}, they still achieve fairly good results. We note that, since we assume the lifetime of $^{259}_{99}$Es to be zero, as described in Table~\ref{tab:info}, the $N_4$ result for the seed nuclide $^{255}_{99}$Es is not shown in Fig.~\ref{fig:sec_abu_prod_N_res12}. However, $^{259}_{99}$Es nuclei are still produced during the interaction. In total a number $N_{4}^{\text{tot}}$ of about $10^3$ of such nuclei is produced at an average neutron energy of $100$ keV, while it decreases to $\sim 1$ as the incident neutron energy goes up to $\sim 1$ MeV.

Furthermore, the maximum values of the abundances $N_i$ ($i$ up to $4$) shown in  Fig.~\ref{fig:sec_abu_prod_N_res12} are found at the end of the $10^4$ neutron pulses, with the only exception being the $^{256}_{99}$Es isotope of $^{255}_{99}$Es. For this case saturation occurs at low and high neutron energy, where $N_1$ already peaks at around $9000$ neutron pulses. This is due to the quite short lifetime of $^{256}_{99}$Es.

We note that $^{126}_{51}$Sb, $^{176}_{71}$Lu and $^{187}_{75}$Re are at or are near the branching point isotopes of the $s$-process \cite{KlayPRC1991, DollPRC1999, KappelerAPJ1993, KappelerAPJ1991, KappelerPPNP1999, BattagliaEPJA2016}. The elements Lu, Re, Os, Tm, Ir and Au are close to the region of the the waiting point $N = 126$ of the $r$-process \cite{PanovAA2009, NegoitaRRP2016}, where $N$ is the neutron number of isotopes. This waiting point is the last point at which the $r$-process path approaches the valley of stability, due to the low binding energy and the high half-life at the shell closure, before the production of the heaviest nuclides. It has been shown in the sensitivity study \cite{MumpowerPPNP2016}, that the abundances of nuclides around this point are highly influenced by the uncertainties of the properties of these nuclides. Measurements of the properties of the neutron-rich nuclei produced from these seed nuclides, as well as the neutron capture cascade itself, would improve our understanding of the neutron-capture nucleosynthesis in astrophysics. In addition, $^{248}_{95}$Am produced by $^{244}_{95}$Am capturing $4$ neutrons, and $^{258}_{99}$Es and $^{259}_{99}$Es produced by $^{255}_{99}$Es capturing $3$ and $4$ neutrons, respectively, are beyond the heaviest isotopes of Am and Es that have been accessed so far in the laboratory \cite{ThoennessenRPP2013, ThoennessenBook2016, ThoennessenIJMPE2014, ThoennessenIJMPE2015, ThoennessenIJMPE2016, ThoennessenIJMPE2017, ThoennessenIJMPE2018, ThoennessenIJMPE2019}. This indicates that we could get access to such neutron-rich isotopes in a regime that has never been access to by other means in the laboratory. Furthermore, $^{129}_{51}$Sb is also an interesting case for fundamental nuclear physics as it has recently been found to show the signal of emerging nuclear collectivity \cite{GrayPRL2020}.

It is important to mention about the accessibility of the reaction products compared with the neutron-rich isotopes produced in conventional accelerator- and reactor-based facilities. Using the intense laser-driven neutron source with relatively lower energy, the neutron-rich isotopes can be produced and accumulate in a small target, leading to a large value for the fraction of the accumulated reaction products in the target. In such a way, the produced neutron-rich isotopes can be easily access after the production. We note that, in conventional accelerator- and reactor-based facilities, as the flux of the beam cannot be that high, it either need very high energy of the beam (e.g., projectile fragmentation, projectile fission or nuclear fusion reactions) or very long time of the accumulation with much thicker target (e.g., neutron capture) for the production of neutron-rich isotopes. With very high energy of the beam, the recoiled energy of the reaction products is also very high that they cannot accumulate in the target, such that one has to measure the reaction products during the production with a lot of other high energy particles. With a very long time of the accumulation with much thicker target for the production of neutron-rich isotopes, the fraction of the accumulated reaction products in the target is small, and the relatively short-lived isotopes cannot stay for the long period of the production process.

In the Petawatt-class laser facilities such as the ELI facilities \cite{NegoitaRRP2016, ELIweb2020} available in the near future, lasers with a power on the Petawatt level and a repetition rate of around $1$ Hz will be in operation. The intense laser-driven neutron beams with $10^{12}$ neutrons per pulse and a high repetition rate that we have mainly focused on in the present work are expected to be achievable in such laser facilities.  In high-power laser facilities currently available, with less power or smaller repetition rate of the lasers than the upcoming ones, the achievable neutron beams are less intense or have a smaller repetition rate compared to the one ($10^{12}$ neutrons per pulse and a repetition of $1$ Hz) that we have mainly focused on \cite{nsrcrit, PomerantzPRL2014, HigginsonPRL2015, nsrcwu, PomerantzELIS2015}. In this case, as shown in Fig.~\ref{fig:best_abu_prod_N_res10}, up to $3$ successive neutron capture events leading to neutron-rich isotopes with $3$ more neutrons than the original seed nuclide are expected.

%%----------------------------------------------------------summary-----------------------------------------------------------

\section{Summary \label{sec:sum}}

We have studied the neutron capture cascades, and consequently, the production of neutron-rich isotopes taking place in single-component targets being irradiated by a laser-driven (pulsed) neutron source. Specifically, we have considered a rectangular target, and investigated the effects of the neutron irradiation for a variety of different target seed nuclides. These seed nuclides have been taken from the recent ENDF-B-VIII.0 neutron sublibrary \cite{ENDFBVIII0}, where we have chosen the heaviest sufficiently-long-lived isotope ($T_{1/2}> 1$ h) per element. In this way a total of $95$ different seed nuclides from $^7_3$Li to $^{255}_{100}$Fm have been studied. Our calculations involve the successive radiative neutron capture process, the damping of the incident neutron beam, the loss of target nuclei by transmutation and radioactive decay of the nuclei, and the effect of multiple neutron pulses. Our calculations show that, even with a single neutron pulse of $10^{12}$ neutrons per pulse, observing $2$ successive neutron capture events is possible. Furthermore, our results for the scenario of $10^4$ neutron pulses, provided by a laser-driven neutron source delivering $10^{12}$ neutrons per pulse at a repetition rate of $1$ Hz, show the possibility of observing up to $4$ successive neutron capture events leading to the production of neutron-rich isotopes with $4$ more neutrons than the original seed nuclide. Such intense laser-driven neutron beams with a high repetition rate are expected to be achievable in the Petawatt-class laser facilities, such as ELI facilities \cite{NegoitaRRP2016, ELIweb2020}, available in the near future. With the neutron beams with less intensity or smaller repetition rate currently available \cite{nsrcrit, PomerantzPRL2014, HigginsonPRL2015, nsrcwu, PomerantzELIS2015}, up to $3$ successive neutron capture events, leading to neutron-rich isotopes with $3$ more neutrons than the original seed nuclide, are expected.

The seed nuclides $^{75}_{33}$As, $^{126}_{51}$Sb, $^{176}_{71}$Lu, $^{187}_{75}$Re, $^{192}_{76}$Os, $^{226}_{88}$Ra, $^{233}_{91}$Pa and $^{244}_{95}$Am, as well as $^{171}_{69}$Tm, $^{193}_{77}$Ir, $^{197}_{79}$Au, $^{227}_{89}$Ac and $^{255}_{99}$Es have been identified as good candidates as they exhibit a good performance and show interesting features. Among these identified interesting seed nuclides, some are in the region of the branching point of the $s$-process ($^{126}_{51}$Sb, $^{176}_{71}$Lu and $^{187}_{75}$Re) or the waiting point of the $r$-process (Lu, Re, Os, Tm, Ir and Au). Moreover it is also possible to produce neutron-rich isotopes ($^{248}_{95}$Am, $^{258}_{99}$Es and $^{259}_{99}$Es) in a regime that has not been accessed by other means in the laboratory. Measuring the properties of the produced neutron-rich nuclei, as well as observing the neutron capture cascades for these seed nuclides which could allow us to simulate the astrophysical neutron capture nucleosynthesis in the laboratory, could improve our understanding of the astrophysical nucleosynthesis. In addition, $^{129}_{51}$Sb produced from the seed nuclide $^{126}_{51}$Sb is an interesting nuclide for fundamental nuclear physics. Our study also shows that the production of neutron-enriched isotopes scales as $N_{i}^{\text{tot}} \propto (N_{pl})^i$ with the number of neutron pulses $N_{pl}$, as long as saturation due to competition among neutron capture, radioactive decay of nuclei, and the loss of nuclei due to transmutation does not occur. Furthermore, the abundance $N_i$ and the total produced amount $N_i^{\text{tot}}$ of neutron-enriched nuclei scale as $N_p^i/A^{i-1}$ before saturation occurs. Our study would be interesting for the industry of radioisotope production, astrophysics concerned with neutron-capture nucleosynthesis, and fundamental nuclear physics.

\begin{acknowledgments}
The authors gratefully acknowledge fruitful discussions with A. P\'alffy, A.-M. Gl\"uck and C. H. Keitel.
\end{acknowledgments}

\appendix

\section{Comparison of NON-SMOKER data and ENDF-B data \label{sec:apddata}}

%%%%%%%%%%%%%%%%%%%%%%%%%%%%%%%%%%%%%%%
\begin{figure*}[!ht]
  \includegraphics[scale=0.95]{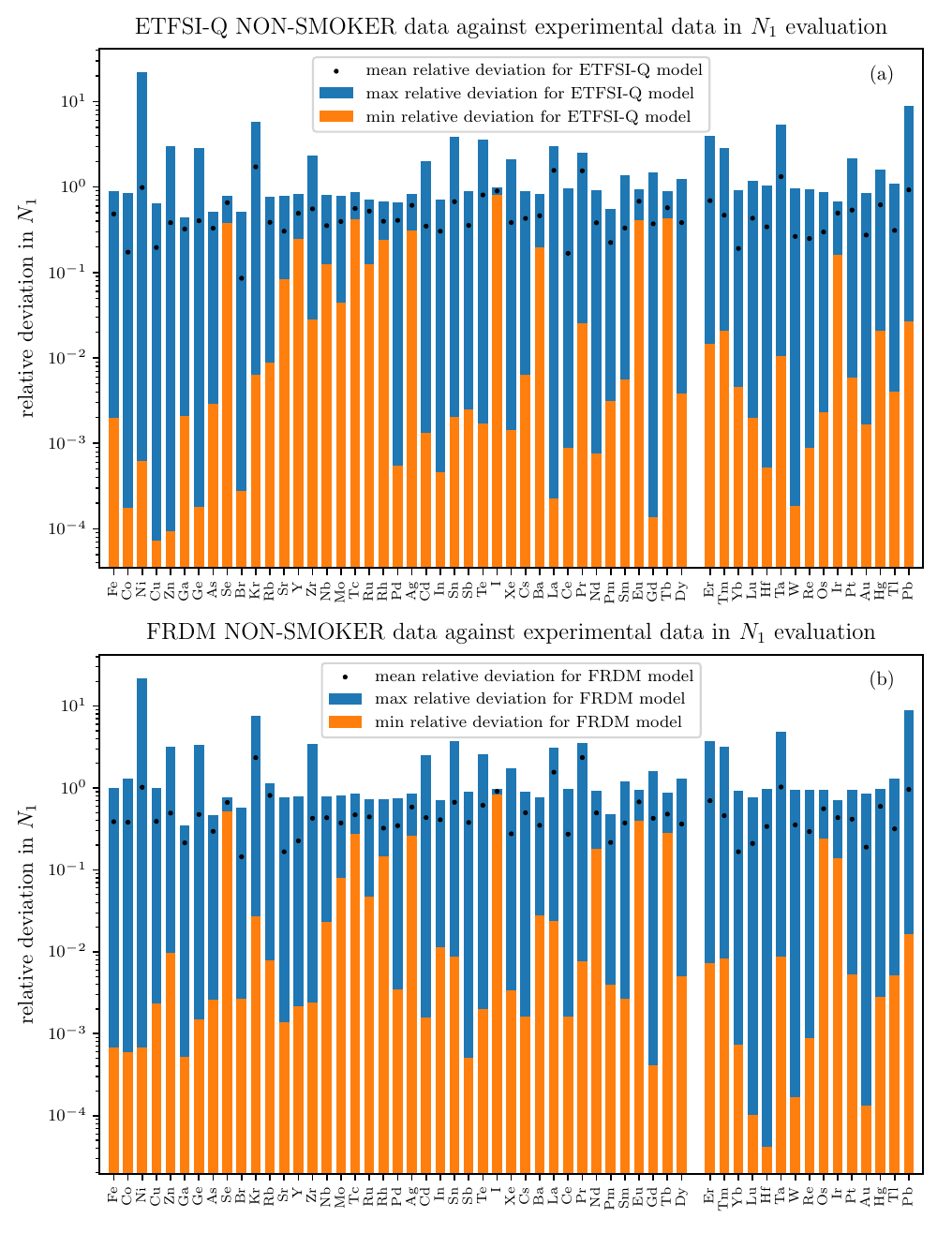}
  \caption{The deviations for $N_1$ after one neutron pulse between the NON-SMOKER data and ENDF-B data. The deviations are obtained by calculating $N_1$ for $300$ log-spaced energies (average incident neutron energy) between $50$ keV to $10$ MeV. A neutron beam with $N_p = 10^{11}$ n/pl is assumed. The damping of the incident neutron beam due to the target thickness, and the radioactive decay of nuclei are not taken into account. (a): Relative deviations between $N_1$ calculated using the NON-SMOKER ETFSI-Q data and $N_1$ calculated using the ENDF-B data. (b): Relative deviations between $N_1$ calculated using the NON-SMOKER FRDM data and $N_1$ using the ENDF data. The blue bar shows the maximal relative deviation and the orange bar shows the minimal relative deviation. The filled circle shows the mean relative deviation for each seed nuclide.}
  \label{fig:theo-exp_comp}
\end{figure*}
%%%%%%%%%%%%%%%%%%%%%%%%%%%%%%%%%%%%%%%

In order to have an idea of the quality of the NON-SMOKER data we employ in the present work, we make quantitative comparisons between the NON-SMOKER data \cite{RAUSCHER200147} and ENDF-B data \cite{ENDFBVIII0}. We therefore calculate $N_1$ by Eq.~(\ref{eq:1pl}) (one neutron pulse, no radioactive decay and no damping of the incident neutron beam) for different seed nuclides and different average neutron energies. Here, we only keep terms up to the leading order for $N_1$. The calculation is performed twice using the two NON-SMOKER datasets and once using the ENDF-B cross section data. We define the relative deviation by
\begin{equation}
  \text{relative deviation} = \left|\frac{N_1|_{\text{NON-SMOKER}} - N_1|_{\text{ENDFB}}}{N_1|_{\text{ENDFB}}}\right|,
\end{equation}
where $N_1|_{\text{NON-SMOKER}}$ denotes $N_1$ calculated using the NON-SMOKER neutron-capture cross sections and $N_1|_{\text{ENDFB}}$ is $N_1$ calculated using the neutron-capture cross sections from the ENDF-B library.

These relative deviations are calculated for $300$ log-spaced energies (average incident neutron energy) between $50$ keV to $10$ MeV. The maximal relative deviation, the minimal relative deviation and the mean relative deviation for each seed nuclide are shown in Fig.~\ref{fig:theo-exp_comp}. Only some of the seed nuclides listed in Table~\ref{tab:info} have been selected for this calculation, since not for all nuclides data is available in all of the three datasets. As in the main text of the paper, we have assumed a target with interacting surface area $A = 25$ $\mu$m and thickness $L = 100$ $\mu$m. The neutron beam has a Gaussian profile energy spectrum with relative width of $w/E_{inc} = 10\%$, and the neutron number per pulse assumed here is $N_p = 10^{11}$ n/pl. Both models used in the NON-SMOKER code yield a mean relative deviation (average over all the seed nuclides presented in Fig.~\ref{fig:theo-exp_comp}) to the ENDF-B data of around $50 \%$, i.e., for the NON-SMOKER ETFSI-Q data, the mean relative deviation is $52\%$, while for the NON-SMOKER FRDM data, the mean relative deviation is $54\%$.

%\clearpage

\bibliography{refsneuc.bib}

\end{document}